\documentclass[12pt]{article}
\usepackage[utf8]{inputenc}
\usepackage[pdftex]{graphicx}   
\usepackage{mathtools}
\usepackage{amsmath}
\usepackage{amssymb}
\usepackage{bbm} 
\usepackage[english]{babel}
\usepackage{amsthm}
\usepackage[sort]{natbib}   
\usepackage{url}
\usepackage{algorithm}
\usepackage{algpseudocode}
\usepackage{algpseudocode}
\usepackage{xcolor}
\usepackage{caption}
\usepackage{subcaption}
\usepackage{placeins}

\usepackage{etoolbox}
\makeatletter
\newenvironment{breakablealgorithm}
{  \refstepcounter{algorithm}
   \par\noindent
   \hrule height.8pt depth0pt \kern2pt
   \textbf{\fname@algorithm~\thealgorithm}\ Approximate Ferguson--Klass Algorithm Using Adaptive Numerical Integration\par
   \hrule\kern2pt
   \begin{center}
}
{
   \end{center}
   \kern2pt\hrule\relax\par
}
\makeatother

\theoremstyle{definition}

\newtheorem{propinner}{Proposition}
\newenvironment{prop}[1]{%
  \IfBlankTF{#1}
    {\renewcommand{\thepropinner}{\unskip}}
    {\renewcommand\thepropinner{#1}}%
  \propinner
}{\endpropinner}

\renewcommand{\d}[1]{\ensuremath{\operatorname{d}\!{#1}}}

\newcommand{\blind}{0}

\begin{document}
\def\spacingset#1{\renewcommand{\baselinestretch}%
{#1}\small\normalsize} \spacingset{1}
\if0\blind
{
  \title{\bf A General Purpose Approximation to the Ferguson-Klass Algorithm for Sampling from L\'evy Processes Without Gaussian Components}
  \author{Dawid Bernaciak 
    \hspace{.2cm}\\
    Department of Statistical Science, University College London.\\
    and \\
    Jim E. Griffin \\
    Department of Statistical Science, University College London}
  \maketitle
} \fi

\if1\blind
{
    \title{\bf A General Purpose Approximation to the Ferguson-Klass Algorithm for Sampling from L\'evy Processes Without Gaussian Components}
  \maketitle
  \medskip
} \fi

\bigskip

\maketitle
\begin{abstract}
    We propose a general-purpose method 
    for generating samples from L\'evy processes without Gaussian components.
    It uses a multi-part approximations of the jump intensity on a grid and 
    applies  the Ferguson-Klass algorithm. We consider how the choice of grid affects the approximation error and propose adaptive selection methods that lead to negligible approximation error. The proposed method is shown to be orders of magnitude faster than the original Ferguson-Klass algorithm and competitive with tailored methods.
 The method opens an avenue for computationally efficient and scalable Bayesian nonparametric models which go beyond conjugacy assumptions, as demonstrated in the examples section.
\end{abstract}

\noindent%
{\it Keywords:}  Completely random measure; Bayesian nonparametrics; gamma process; beta process; generalized gamma process; stable-beta process
\vfill

\section{Introduction}

Completely random measures (CRMs) \citep{kingman1967completely} serve as a fundamental component of Bayesian nonparametric statistics and 
  have been utilised to establish tractable priors for random distributions. Their applications include density estimation \citep{regazzini2003distributional} and, clustering via the Chinese Restaurant Process \citep{aldous1985exchangeability}. The latter is 
 based on the Dirichlet process \citep{ferguson1973bayesian,sethuraman1994constructive} which is also used as the mixing measure in the popular Dirichlet process mixture model \citep{antoniak1974mixtures} for density estimation. The Dirichlet process can be represented as the normalisation of gamma process, which is a CRM.
In addition, CRMs are used in matrix factorisation through the Indian Buffet Process \citep{ghahramani2005infinite}, which builds on the beta process \citep{hjort1990nonparametric} as shown by \cite{thibaux2007hierarchical}. 
Other notable CRMs that are utilized for their power-law characteristics include the stable-beta process \citep{teh2009indian}, the generalised gamma process \citep{hougaard1986survival,brix1999generalized}, and the Pitman-Yor process \citep{pitman1995exchangeable}, which can be expressed either as a mixture of normalised generalised gamma processes or a normalised $\sigma$-stable process \citep{pitman1997two}.
Recent advances in the field have introduced processes with double power-law behaviour such as the generalised BRFY process \citep{ayed2019beyond} which provide a wide range of modelling choices.
Usually, the sub-class  of CRMs
without fixed points of discontinuity is used. These have an infinite number of jumps at random locations and 
can be characterised through their L\'evy intensity in a L\'evy-Khintchine representation \citep{DalVJ03}.
\cite{ghosal2017fundamentals} provide a comprehensive introduction to completely random measures and random processes.

Realisations of CRMs are usually simulated by truncating a particular series representation. 
\cite{campbell2019truncated} present a comprehensive review of series representation methods and their truncation errors.
The representations include 
inverse L\'evy process or Ferguson-Klass representation \citep{ferguson1972representation},
Bondesson representation \citep{bondesson1982simulation}, thinning representation \citep{Ros90}, rejection representation \citep{rosinski2001series}, and
stick-breaking representation \citep{sethuraman1994constructive, 
paisley2010stick, Fav16}. These can have convenient forms. For example, the stick-breaking representation of the beta process \citep{paisley2010stick} 
involves only beta random variables. However, in general, they can involve non-standard probability distributions, or functions with no analytical inverse and/or integral.

In Bayesian nonparametric inference,
simulation-based inference is challenging
\citep{griffin2019two, nguyen2023independent} with CRMs 
outside the standard families
 arising as either posterior distributions or full conditional distributions in Markov chain Monte Carlo samplers. Addressing this challenge has lead to the development of methods which are limited to specific processes \citep{teh2007stick, teh2010hierarchical,griffin2011posterior,regazzini2003distributional}, which
are flexible but computationally demanding \citep{lomeli2017marginal,favaro2013slice,favaro2012new}, or which allow  the CRM  to be marginalized from the posterior
 \citep[e.g.][]{EW1995, Neal2000, ghahramani2005infinite, favaro2013mcmc}.  
In practice, the lack of a fast, general-purpose methods has lead to  models  that either  conjugate or rely on specifically tailored techniques  to circumvent the problem of sampling from a non-standard CRM.

This gap in the literature
motivates the development of a computationally efficient general-purpose 
method for  L\'evy process without Gaussian components which correspond to CRMs in one-dimensions (specifically, 
 subordinators which are non-negative pure jump L\'evy process in one dimension). 
We build on the simple and general Ferguson-Klass algorithm. Its truncation 
error properties  \citep{arbel2017moment, campbell2019truncated} are well understood
and lead to the smallest average truncation error for a fixed number of jumps
(\cite{arbel2017moment} and
\cite{zhang2024posterior} propose a way to remediate the error).
Our algorithm\footnote{Our methodology is implemented as an open source Python package available at 
\url{https://github.com/dbernaciak/bayes-crm}. 
The website also reproduces all results presented in this paper.}
uses an approximation to the L\'evy intensity 
that avoids  repeated (numerical) inversions often needed in implementation of the Ferguson-Klass algorithm
and can be used as a ``black box'' method for sampling
 from CRMs.
We compare performance to standard methods for commonly used CRMs and applications to two examples with non-standard CRMs: sampling from the full conditional distribution in 
a compound random measures \citep{griffin2017compound}
and the posterior distribution in a species sampling model with observational errors. These
 demonstrate the computational simplicity and scalability
 of our sampling 
approach, which can unlock future development and use of Bayesian nonparametric methods.

The paper is organised as follows. Section \ref{sec:background} provides a brief theoretical introduction to completely random measures and the Ferguson-Klass algorithm. Section \ref{sec:methodology} outlines the proposed methodology and Section \ref{sec:numerical} presents the empirical results of the precision analysis and the performance benchmarks for the standard CRMs. Section \ref{sec:examples} displays applications of the proposed method. First, in the context of compound random measures, we benchmark the speed performance of our method against the standard Ferguson-Klass. The second example illustrates the use of our method in the species-sampling problem where we construct a simple but non-conjugate Bayesian nonparametric model and show how our method facilitates the ease of computation, which was unachievable before.
\section{Background}
\label{sec:background}

Completely random measures (CRMs) without fixed points of discontinuity \citep{kingman1967completely} on some general measure space $B$ can be represented
as a measure with an infinite number of jumps in the form 
\begin{equation*}
    G = \sum_{k=1}^{\infty} J_k\, \delta_{\theta_k}
\end{equation*}
where $\delta_x$ is the Dirac delta function which places mass $1$ at $x$, 
$(J_1, \theta_1),(J_2, \theta_2), (J_3, \theta_3), \ldots$ are random, with
$J_1, J_2, J_3, \ldots\in \mathbb{R}_+$ called jump sizes and $\theta_1, \theta_2, \theta_2, \ldots \in B$ called jump locations. The measure $G$ can be characterized using the L\'evy-Khintchine representation 
by
\[
\mbox{E}\left[\exp\left\{-\int_B f(\theta) \, G(d\theta)\right\}\right]
= \exp\left\{-\int_0^{\infty} \int_B (1 - \exp\{- s f(\theta))
\, \bar\nu(ds, d\theta)\right\}
\]
where $f: B\rightarrow \mathbb{R}_+$ is a measureable function for which, almost surely, $\int_B f(\theta) \, dG(\theta) < \infty$ and $\bar{\nu}(\d J, \d \theta)$ is a measure for which
$
\int_{\mathbb{R}^+} \int_A \min\{1, s\} \, \bar\nu(ds, dx) < \infty,
$
for any measureable $A \subset B$. We will refer to $\bar{\nu}(\d J, \d \theta)$ as the 
the  L\'evy intensity and assume that it is homogeneous, i.e., $\bar{\nu}(\d J, \d \theta)=\nu(\d J)\, \alpha(\d \theta)$, where $\nu$ is a measure on $\mathbb{R}^+$ 
for which
$
\int_{\mathbb{R}^+} \min\{1, s\} \, \nu(ds) < \infty,
$
called the jump intensity,
and $\alpha $ is a non-atomic probability measure on $B$. This implies that the jump locations $\theta_1, \theta_2, \theta_3, \dots \stackrel{i.i.d.}{\sim} \alpha$ and we concentrate on simulating the jump sizes for the rest of this paper.
The jump intensities of popular CRMs are presented in Table \ref{tab:crm} together with their parametrisations, such that the total mass $M = \mathbb{E}[\sum_{k=1}^{\infty} J_k]$. 
\begin{table}[h!]
\centering
\begin{tabular}{|l|l|l|l|}
\hline
\textbf{Process} & \boldmath$\nu(x)$ & \textbf{Domain} & \textbf{Parameter range} \\
\hline
Gamma & $Mx^{-1}\exp\{-x\}$ & $x > 0$ & $M > 0$ \\
$\sigma$-stable & $\frac{\sigma}{\Gamma(1-\sigma)} x^{-1-\sigma}$ & $x > 0$ & $0 < \sigma < 1$ \\
Beta & $M cx^{-1}(1-x)^{c-1}$ & $0 < x < 1$ & $c > 0$ \\
Generalized gamma & $M \frac{a^{1-\sigma}}{\Gamma(1-\sigma)} x^{-1-\sigma} \exp\{-ax\}$ & $x > 0$ & $M > 0, 0 < \sigma < 1, a > 0$ \\
Stable-Beta & $M \frac{\Gamma(1+c)}{\Gamma(1-\sigma)\Gamma(c+\sigma)} x^{-1-\sigma} (1-x)^{c+\sigma-1}$ & $0 < x < 1$ & $0 < \sigma < 1, c > 0$ \\
\hline
\end{tabular}
\caption{The jump intensity for various CRM's}
\label{tab:crm}
\end{table}

 Let
us denote the arrival times of a unit rate Poisson process on $\mathbb{R^+}$ as $E_k = \sum_{l=1}^k T_l$ with $T_l \stackrel{i.i.d.}{\sim} \text{Exp}(1)$. 
The tail mass function $\eta_{\nu}$ of a jump intensity $\nu$ is defined as $\eta_{\nu}(x) = \int_{x}^{\infty} \nu(z) \d z$, which is a measure on $\mathbb{R_+}$ and the inverse tail measure $\eta_{\nu}^{-1}(z) \coloneqq \inf \left\{ x: \eta_{\nu}(x) \leq z \right\}$. $G$ has an \textit{inverse-L\'evy} representation, denoted $G \leftarrow \text{IL-Rep}(\nu)$ if
\begin{equation*}
    G = \sum_{k=1}^{\infty} J_k \,\delta_{\theta_k}, \qquad J_k = \eta_{\nu}^{-1}(E_k).
\end{equation*}
\cite{ferguson1972representation} proved that $G \leftarrow \text{IL-Rep}(\nu)$ implies $G\sim \text{CRM}(\nu)$. This observation led to a widely used algorithm for simulation of L\'evy processes without Gaussian components which is often called the Ferguson-Klass algorithm and has the steps (a  fuller pseudo-code version is provided in Appendix \ref{alg:ferguson-klass}). Set $E_1 \sim \mbox{Exp}(1)$ and $i = 1$
\begin{enumerate}
\item Set $J_k = \eta_{\nu}^{-1}(E_k)$,
\item Set $E_{k + 1} = E_k + T_k$ where $T_k\sim\mbox{Exp}(1)$ and $k = k + 1$,
\item Return to step 1.
\end{enumerate}
This produces the jump sizes in an almost surely decreasing order, i.e., $J_1 \geq J_2 \geq J_3 \geq \ldots$ and is stopped either when $i = N$ or $J_i < \epsilon$ for user-chosen values of $N$ or $\epsilon$.  This approach can be applied to general $\nu$, but in practice it can be slow since  the tail mass function has no analytical solution for many popular CRMs  and so the tail mass function $\eta_{\nu}$ needs to be numerically integrated and inverted multiple times.

\section{Methodology}
\label{sec:methodology}

We develop a multi-part approximation $\tilde\nu$ of the jump intensity $\nu$,
  which
allows the inverse of the tail mass function $\eta_{\tilde\nu}^{-1}$ to be efficiently computed. This can be used as a numerical quadrature method for tail masses of L\'evy intensities. If the multi-part approximation is also an envelope
({\it i.e.} $\tilde\nu(x) > \nu(x)$ for all $x\in B$) then  a rejection sampling algorithm \citep{rosinski2001series} with 
negligible expected probability of rejection/thinning can be defined.
The thinning breaks the one-to-one correspondence between the input arrival times and the output jump sizes, which might not be desirable for some computational problems. We discuss this algorithm and applications to jump intensities commonly used in Bayesian nonparametrics in Section \ref{sec:examples}. All proofs are grouped in 
 Appendix~\ref{appendix:integration} and
 Appendix~\ref{sec:grid-selection}.

The multi-part approximation is defined by specifying a grid $x_0< \dots< x_n$ ($n+1$ points which specify $n$ bins), and a piecewise approximation $\tilde{\nu}(x)$
 \begin{equation}
 \tilde{\nu}(x) = f_i(x),\qquad
 x_{i-1} < x \leq x_i, \quad i = 1,\dots, n
 \label{approx_f}
 \end{equation}
where $f_i(x) > 0$ approximates $\nu(x)$ for $x_{i-1} <x \leq x_i$.

\subsection{Defining $f_i$}
\label{section:fi}

\begin{figure}[ht]
    \centering
    \begin{subfigure}{0.61\textwidth}
    \centering
    \includegraphics[width=1\textwidth]{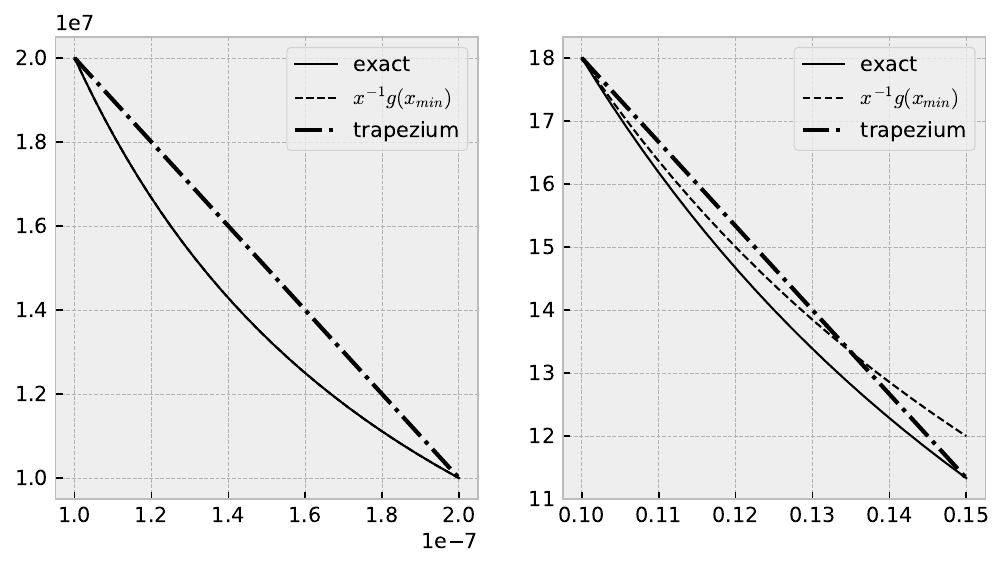}
    \caption{Numerical integration method illustration for the Beta process with $M=1$ and $c=2$. On the left, illustration for the interval $x \in (2\times10^{-7}, 1\times10^{-7}]$ and on the right illustration for the interval $x \in (0.15, 0.1]$. The $x^{-\kappa}$ approximation provides a better fit for $x$ than the the trapezoidal method for small \(x\).}
    \label{fig:integration}
    \end{subfigure}
    \hfill
     \begin{subfigure}{0.35\textwidth}
        \centering
        \includegraphics[width=\textwidth]{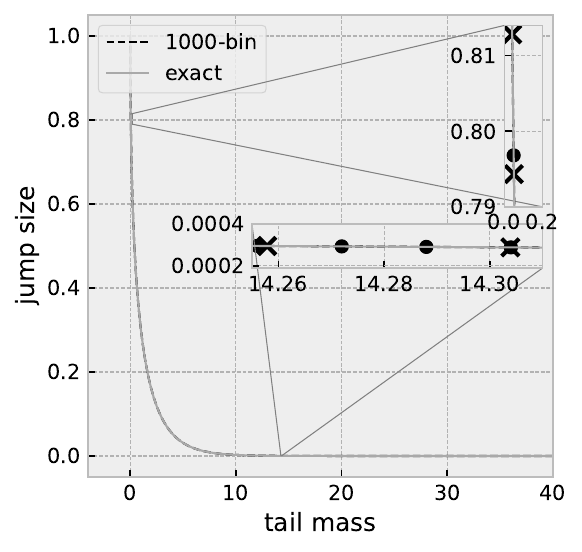}
        \caption{Algorithm illustration for the Beta process with $M=1$ and $c=2$. The crosses represent points that were not interpolated, the dots represent the interpolated values.}
        \label{fig:sub1}
    \end{subfigure}
    \caption{}
    \label{fig:methodology}
\end{figure}

In general, we use piece-wise linear functions
\[
f_i(x) = 
\nu(x_{i-1}) + \frac{(\nu(x_{i-1}) - \nu(x_{i}))}{x_{i} - x_{i-1}}(x_{i} - x), \qquad  x_{i-1} \leq x < x_{i}, 
\]
which leads to a trapezoidal integration rule. Alternatively, if $\nu(x)$ is approximately polynomial near zero so that
$\nu(x) = h(x)\, g(x)$
on some interval   $( 0, x_{\mathrm{thr}}]$, where $h(x) = x^{- \kappa}$ and 
$g(x)$ is differential and bounded, 
we use
\[
f_i(x) =
\begin{cases}
\nu(x_{i-1})
\left(\frac{x}{x_{i-1}}\right)^{-\kappa}
, \qquad & x_{i-1} \leq x < x_i \text{ and } x_{i-1} < x_{\mathrm{thr}}, \\[1mm]
\nu(x_{i-1}) + \dfrac{\nu(x_{i-1})-\nu(x_{i})}{x_{i}-x_{i-1}}\,\bigl(x_{i}-x\bigr),\qquad  & x_{i-1} \leq x < x_{i} \text{ and } x_{i-1} \ge x_{\mathrm{thr}}.
\end{cases}.
\]
This decomposition is available for well-known processes such as beta, gamma, or generalised gamma processes.
 Both methods are illustrated in Fig.\ \ref{fig:integration}. We consider these choices for two reasons. First, the resulting approximation 
  $\tilde{\nu}(x)$ is an envelope of $\nu(x)$  for the intensities used in Bayesian nonparametrics, which allows us to correct the approximation error by the rejection algorithm. Second, this integration quadrature is computationally easy and efficient. For example, using a more sophisticated quadrature like Simpson's rule would lead to better convergence, but the computation of error would be much more complex to calculate and the error harder to correct via thinning due to the quadratic shape of the approximation.

Using the trapezoidal rule, the mass on the $i$-th bin, $b_i = \int_{x_{i-1}}^{x_i} f_{i}(x) \, dx$, is 
\begin{equation*}
    b_i = \int_{x_{i-1}}^{x_{i}} \tilde\nu(z) \, dz = \frac{(\nu(x_{i}) + \nu(x_{i-1})) (x_{i} - x_{i-1})}{2}
\end{equation*}
and,
with the polynomial approximation,
in the region \([x_0,x_{\mathrm{thr}}]\), the mass of the $i$-th bin is 
\[
b_i = \int_{x_{i-1}}^{x_{i}} \nu(z)\,dz =
\frac{g(x_{i-1})}{1- \kappa}
\left(\frac{x_i}{x_{i-1}} x_i^{-\kappa} - x_{i-1}^{-\kappa}\right).
\]

Applying the Ferguson-Klass algorithm to $\tilde\nu$, 
we find the index $i$ such that \(\eta_{\tilde{\nu}}(x_{i}) \leq E_k \leq \eta_{\tilde{\nu}}(x_{i-1})\),
where \(\eta_{\tilde{\nu}}(x_{i}) = \sum_{j=i}^{n-1} b_j \), and calculate
\[
     \tilde{J}_k= x_{i} + \delta(E_k, i),
\]
where $\delta(E_k, i)$ is calculated either by inversion of \(f_i(x)\) or, for improved execution speed, linear approximation to the inversion 
\[
\delta(E_k, i) = \frac{x_{i-1} - x_{i}}{\eta_{\tilde{\nu}}(x_{i-1}) - \eta_{\tilde{\nu}}(x_{i})} \left(E_k - \eta_{\tilde{\nu}}(x_{i})\right).
\]

\subsection{Error analysis}
\label{section:error}

We define $\tilde{G}$ to be a L\'evy process
with 
L\'evy intensity $\tilde{\nu}(\d J)\, \alpha(\d \theta)$ and define $\tilde{J}_1, \tilde{J}_2,  \tilde{J}_3, \ldots$ to be the jumps  of $\tilde{G}$ simulated using the Ferguson-Klass algorithm. Let $C^+ =\{x \in B\mid \tilde\nu(x) \geq \nu(x)\}$ and
$C^- =\{x \in B\mid \tilde\nu(x) < \nu(x)\}$, and
$\phi^+$  and $\phi^{-}$ be L\'evy processes with L\'evy intensities 
$\left(\tilde{\nu}(dx) - \nu(dx)\right) \, \alpha(d\theta)\,\mbox{I}(x \in C^+)$
and $\left(\nu(dx) - \tilde\nu(dx)\right) \, \alpha(d\theta)\,\mbox{I}(x \in C^-)$ respectively.
Clearly, we can write
\[
\tilde{G} = G + \phi^+ - \phi^-
\]
where 
 the number of points in $\phi^+$ and $\phi^{-}$ are independent and  Poisson distributed with means $\int_0^{\infty} \mbox{I}(x \in C^+) \left[ \tilde{\nu}(x) - \nu(x) \right] d x$
and
$\int_0^{\infty} \mbox{I}(x \in C^-)\left[ \nu(x) - \tilde\nu(x) \right] d x$ respectively. The bias of the approximation on a measurable set $A\subset B$ is
\[
\mbox{E}[\tilde{G} - G](A) =
\mbox{E}[\phi^+](A) - 
\mbox{E}[\phi^-](A)
= 
\int x\left(\tilde{\nu}(x) - \nu(x)\right) \,dx \, \alpha(A).
\]
This characterizes how the difference between the L\'evy intensity and the approximation affects the total mass. 
Since $\tilde\nu$ is only defined on $(x_0, x_n)$,
 there are three sources of error:
\[
\mbox{E}[\tilde{G} - G](A) = 
\left\{
\overbrace{\int_{x_0}^{x_n} x\left(\tilde{\nu}(x) - \nu(x)\right) \,dx}^{\mbox{approximation error}}
- \overbrace{\int_0^{x_0} x\, \nu(x)\,dx}^{\mbox{small jump error}}
- \overbrace{\int_{x_n}^{\infty} x\, \nu(x)\,dx}^{\mbox{large jump error}}
\right\} \,\alpha(A).
\]
which are 
the approximation error of using $\tilde\nu$ rather than $\nu$, and truncation errors due to the absence of small jumps ({\it small jump error}) and the absence of large jumps ({\it large jump error}). 

In practice,
 the Ferguson-Klass algorithm is stopped at some point to simulate a truncated version of $G$. We define the processes 
$G^{(\epsilon)}$ and $\tilde{G}^{(\epsilon)}$ by  removing jumps smaller than $\epsilon$ from $G$ and $\tilde{G}$ respectively. Choosing $x_0 = \epsilon$ leads to 
\[
\mbox{E}[\tilde{G}^{(x_0)} - G^{(x_0)}](A) = 
\left\{
\overbrace{\int_{x_0}^{x_n} x\left(\tilde{\nu}(x) - \nu(x)\right) \,dx}^{\mbox{approximation error}}
- \overbrace{\int_{x_n}^{\infty} x\, \nu(x)\,dx}^{\mbox{large jump error}}
\right\} \,\alpha(A).
\]
Similarly, let
$G^{(N)}$ and $\tilde{G}^{(N)}$
be $G$ and $\tilde{G}$ 
truncated to the $N$ largest  jumps.
We simulate $E_1, \dots, E_N$ from a unit rate Poisson process and choose $x_0 < \eta_{\tilde\nu}^{-1}(E_N)$ 
(see Section~\ref{section:lower} for details). This leads to 
\[
\mbox{E}[\tilde{G}^{(N)} - G^{(N)}](A) = 
\left\{
\overbrace{\int_{\max\{J_N,\tilde{J}_N\}}^{x_n} \left(\tilde\nu(x) - \nu(x)\right) \, dx 
+
\Delta
}^{\mbox{approximation error}}
- \overbrace{\int_{x_n}^{\infty} x\, \nu(x)\,dx}^{\mbox{large jump error}}
\right\} \, \alpha(A).
\]
where $J_N = \eta_{\nu}^{-1}(E_N)$, $\tilde{J}_N = \eta_{\tilde\nu}^{-1}(E_N)$
and 
\[
\Delta = \int_{\min\{J_N, \tilde{J}_N\}}^{\max\{J_N, \tilde{J}_N\}} 
x\,
\left(\mbox{I}(J_N > \tilde{J}_N)
 \,\tilde\nu(x) 
 -\mbox{I}(J_N \leq \tilde{J}_N)\,
 \nu(x)\right)
 \,dx.
 \]

We concentrate on the approximation error and the large jump errors in these truncations.
The approximation error is controlled by the choice of $f_1(x), \dots, f_n(x)$, and the grid $x_0, \dots, x_n$. The large jump error by the choice of $x_n$.

Let us define \(I(i) = \int_{x_i}^{x_{i+1}} \nu(x) \d x\),   \(I_{\mathrm{trap}}(i) = \frac{x_{i}+x_{i+1}}{2}[f_i(x)+f_{i+1}(x)]\) and \(I_{\mathrm{poly}}(i) = \frac{\nu(x_i)}{x_i^{-\kappa}} \int_{x_i}^{x_{i+1} }x^{-\kappa} \d x\). Further, we define the integration errors \(e_{\mathrm{trap}}(i) = I(i)-I_{\mathrm{trap}}(i)\) and \(e_{\mathrm{poly}}(i) = I(i)-I_{\mathrm{poly}}(i)\).

\begin{prop}{1}
Suppose that $\nu(x) = h(x)\, g(x)$
on some interval   $( 0, x_{\mathrm{thr}}]$, where $h(x) = x^{- \kappa}$ and 
$g(x)$ is twice differential and bounded, with $g'(x) < -\kappa x^{-1} g(x)$ for all $x\in (0, x_{\mathrm{thr}})$.
Then
$
e_{\mathrm{poly}}(i)=O\Bigl(g'(x_i)h(x_i)(\Delta x_i)^2\Bigr)$,  $e_{\mathrm{trap}}(i)=O\Bigl(g(x_i)h'(x_i)(\Delta x_i)^2\Bigr)$
and
\[
| e_{\mathrm{poly}}(i)| < | e_{\mathrm{trap}}(i)|.
\]
\end{prop}
There exist regions, however, where the trapezoidal method yields a smaller error, as described in the below proposition.
\begin{prop}{2}
There exists a threshold \(x_{\mathrm{thr}}\in[a,b]\) 
and $\epsilon > 0$
such that, for \(\Delta x_i < \epsilon\),
\[
| e_{\mathrm{trap}}(i)| < | e_{\mathrm{poly}}(i)|
\mbox{ for all } x_i\in [x_{\mathrm{thr}},b].
\]
 
\end{prop}
These results support using the polynomial approximation close to zero and the trapezoidal rule away from zero.

The overall error induced by the numerical integration can be quantified. For the trapezoidal rule, the error is given by
\begin{equation*}
    \int_0^{\infty} \tilde{\nu}(x) - \nu(x) \d x = \sum_{i=0}^{n-1} \int_{x_i}^{x_{i+1}} \left(\tilde{\nu}(x_i) - \nu(x)\right) + \frac{x-x_i}{x_{i+1} - x_{i}} \left(\tilde{\nu}(x_{i+1}) - \tilde{\nu}(x_i) \right) \d x.
\end{equation*}
Similarly, when employing the polynomial approximation, the error is
\begin{equation*}
    \int_0^{\infty} \tilde{\nu}(x) - \nu(x) \d x = \sum_{i=0}^{n-1} \int_{x_i}^{x_{i+1}} \left(g(x_i) x^{-\kappa} - \nu(x)\right) \d x.
\end{equation*}
Since the integration accumulates the errors from countably many bins, the order of magnitude of errors is preserved.

We can also estimate the relative error of the jump sizes \(\frac{|J_k-\tilde{J}_k|}{J_k}\). For both the trapezoidal and the polynomial rule, the error in each bin is of order \(e_i=O((\Delta x_i)^2)\). 
For the proposed geometric grid, the relative error magnitude is (as shown in Appendix \ref{appendix:integration})
\begin{equation}
    \frac{|J_k-\tilde{J}_k|}{J_k} \approx O\left((c-1)^2\right).
    \label{eq:relative-error}
\end{equation}

The approximation error can  be removed if $\tilde\nu$ is an envelope for $\nu$, {\it i.e.} if $\tilde\nu(x) > \nu(x)$ for all $x\in B$, which is possible if $f_i(x) \geq \nu(x)$ for $x \in (x_{i-1}, x_i]$. This allows us to define a 
rejection algorithm \citep{rosinski2001series} where $G$ is represented as 
\begin{equation*}
    G = \sum_{k=1}^{\infty} S_k \tilde{J}_k \delta_{\tilde{\theta}_k}.
\end{equation*}
where 
 $S_1, S_2, S_3,\ldots$ are independent Bernoulli random variables with success probability $\frac{\nu(\tilde{J}_k)}{\tilde\nu(\tilde{J}_k)}$. This leads to the algorithm
 \begin{enumerate}
 \item Set $i=1$ and $k=1$.
     \item Simulate $L_k$ with L\'evy intensity $\tilde\nu$.
     \item Simulate $S_k$ as a Bernoulli random variable with success probability $\frac{\nu(\tilde{J}_k)}{\tilde\nu(\tilde{J}_k)}$.
     \item If $S_k=1$, set $J_i = L_k$, $i = i + 1$ and $k = k + 1$. Otherwise, set $k = k + 1$.
     \item Return to step 2.
 \end{enumerate}
The number of  points removed in step 4 is $\sum_{k=1}^{\infty} S_k$ which is Poisson distributed with mean $\mu = \int_0^{\infty} \left[ \tilde{\nu}(x) - \nu(x) \right] \d x$ \citep{campbell2019truncated}. 

Unlike fixed-boundary grid approaches (e.g., the strip method in \citet{Devr86}), our adaptive method dynamically adjusts grid boundaries. This flexibility reduces both approximation and truncation errors by avoiding a priori fixed partitions and large truncation errors. Moreover, it ensures ordered, non-increasing jumps, thereby enhancing computational efficiency \citep{campbell2019truncated}.

\subsection{Grid spacing}
\label{sec:grid-spacing}

Let $x_1, \dots, x_{n-1}$ be grid points defined by
\[
x_i = c^i x_0 ,\qquad i = 1, \dots, n.
\]
for some constant $c>1$. The following propositions specify the optimal grid spacing for a class of intensities used in Bayesian nonparametrics literature. Their proofs can be found in Appendix \ref{sec:grid-selection}.

\begin{prop}{3}
    Suppose the L\'evy intensity $\nu(x)=h(x)g(x)$, $|h'(x)|>|g'(x)|$ on $(0,x_{\mathrm{thr}}]$ and $h(x)=x^{-1}$. Then under the trapezoidal rule  the optimal grid spacing that equidistributes the error satisfies $\Delta x \propto x$. I.e., the geometric grid $x_i = c^i x_0$ is optimal.

\label{prop:geo}
\end{prop}

\begin{prop}{4}
    Suppose the L\'evy intensity $\nu(x)=h(x)g(x)$, $|h'(x)|>|g'(x)|$ on $(0,x_{\mathrm{thr}}]$ and $h(x)=x^{-\kappa}$ with $\kappa \neq 1$. Then under the trapezoidal rule the optimal grid spacing that equidistributes the error satisfies $\Delta x \propto x^{(\kappa+2)/3}$.

\label{prop:geo2}
\end{prop}
We use geometric spacing as the default in this work, but one could incorporate more sophisticated spacing schemes to further minimise the integration error depending on the form of the L\'evy intensity.
In practice, however, for all L\'evy intensities in Table \ref{tab:crm} \( 1\leq \kappa < 2\) so that 
\( 1 \leq (\kappa+2)/3 < 4/3\). Therefore, the geometric grid provides a good approximation to the optimal grid for the values of $\kappa$ used in Bayesian nonparametrics.

The number of grid points affects both the relative error (Section \ref{sec:precision}) and the speed of the execution of the algorithm (Section \ref{sec:performance}). In terms of speed, the  complexity of the proposed algorithm is \(O(n + N)\) and the Ferguson-Klass (Appendix \ref{alg:ferguson-klass}) has complexity \(O(n \times N \times R)\) (assuming that the integration is done on the same number of grid points as in our algorithm), where  $R$ is the average number of iterations in the root finding method. From this analysis, since usually $n \gg N$, we can deduce that our method is around $N \times R$ times faster than the standard implementation of the Ferguson-Klass method.\\

 The rest of the methodology section is organised as follows. In Section \ref{sec:adaptive-grid} we explain the choice of 
 grid  $x_0, x_1, \ldots, x_n$  which adjusts a 
 default choice. Section \ref{section:tail} describes the adjustment to the right to control the large jump error and Section \ref{section:lower}
 describes the search for \(x_0\) which eliminates the small jump error in the simulated realisation. The algorithmic summary of the methodology is presented in Appendix~\ref{sec:algo}. 

\subsection{Adaptive grid specification}
\label{sec:adaptive-grid}

\subsubsection{Adaptive step for $x_n$}
\label{section:tail}

The function $\nu$ is either defined on a bounded interval $(0, b)$ or $\mathbb{R}^+$.
 For bounded domains, we simply set 
$x_n = b$,
which ensures that the large jump error is zero.
 When the domain is \(\mathbb{R}^+\), 
 we choose \(x_n\) as a cutoff point such that
$
\int_{x_n}^{\infty} \nu(z)\,dz < \epsilon_{\text{large}}$,
where $\epsilon_{\text{large}}>0$ is a prescribed tolerance (with the default value $\epsilon_{\text{large}} = 10^{-10}$), computed via the Fortran library QUADPACK and so the large jump error is less than $\epsilon_{\text{large}}$.
To determine the cutoff point in general, we perform a grid search over a geometrically spaced set of candidate cutoff points and define
\begin{equation}
x_n = \min\left\{ x \in \{10^0,\,10^{0.5},\,10^1,\,\ldots\} : \int_{x}^{\infty} \nu(z)\,dz < \epsilon_{\text{large}} \right\}.
\label{gen:xn}
\end{equation}

 Most processes used in Bayesian nonparametrics (and all in Table \ref{tab:crm}) have 
either an  exponentially or polynomially decaying tail mass function. In these cases, $x_n$ can be found without the iterative procedure in Eq.\
\eqref{gen:xn}. The approach can be extended to other tail behaviours, such as double exponential decay. In both cases, we choose an initial value for $x_n$ which is denoted $x_{\mathrm{tail}}$ and find an approximation to $x_n$ for which $\int_{x_n}^{\infty} \nu(z)\, dz < \epsilon_{large}$.

\paragraph{Exponentially decaying tail}
If the tail of the CRM decays as $\exp(-ax)$ (e.g., gamma process), where $a>0$ is some constant. For any $\Delta x > 0$, the ratio 
\begin{align}
    \frac{\int_{x_{\mathrm{tail}}-\Delta x}^{x_{\mathrm{tail}}} \exp(-a z) \d z}{\int_{x_{\mathrm{tail}}}^{x_{\mathrm{tail}}+\Delta x} \exp(-a z) \d z} = \frac{\exp(a \Delta x) - 1}{1 - \exp(-a \Delta x)} = 1/r(a, \Delta x),
\label{eq:exp-dec}
\end{align}
 is independent of $x$. We choose $\Delta x = x_{\mathrm{tail}}(c-1)$. We note that by Eq.\ \eqref{eq:exp-dec} the bin areas form a geometric series \(b, r b, \ldots, r^{n-1} b, r^n b, \ldots\), where \(b=1-\exp(-a \Delta x)\). We need to find \(n\) s.t., \(\epsilon_{\mathrm{large}} = \frac{r^n b}{1-r}\), which gives \(n = \log_r(\frac{\epsilon_{\mathrm{large}}(1-r)}{1-\exp(-a \Delta x)})\). We then set
\[
x_{n} \approx x_{\text{tail}} + \log_{r(a,\Delta x)}\!\Biggl(\frac{\epsilon_{\mathrm{large}}\,(1-r(a,\Delta x))}{1-\exp(-a\Delta x)}\Biggr) \Delta x.
\]

\paragraph{Polynomially decaying tail}
\label{sec:poly-tail}

If the tail is polynomially decaying, 
{\it i.e.} $x^{-p}$ with $1 \neq p > 0$ (e.g., like a $\sigma$-stable process). This may be known from the form of L\'evy density or it can be estimated using the following method.
For $c > 1$,
\begin{align}
K = \frac{\int_{cx}^{c^2x} z^{-p} \d z}{\int_{x}^{cx} z^{-p} \d z} = c^{1-p},
\label{eq:p-decay}
\end{align}
 is independent of $x$. To find $p$, 
we approximate the integrals in 
 \eqref{eq:p-decay}
  by trapezoids, $K\approx\frac{c(\nu(c^2x)+\nu(cx))}{\nu(cx)+\nu(x)}$
 and use   $p = 1 - \log_c(K)$. The tail integral can be approximated as \(\eta_{\nu}(x) \approx  \frac{\nu(x_{\mathrm{tail}})}{x_{\mathrm{tail}}^{-p}}\int_{x}^{\infty} z^{-p} \d z\) which implies that 
 \[
 x_{n} \approx \left(\frac{\epsilon_{\mathrm{large}} \, p \, x_{\mathrm{tail}}^{-p}}{\nu(x_{\mathrm{tail}})}\right)^{1/(1-p)}.
 \]
 
For polynomially decaying tails, the advantage of a geometrically spaced grid in this case is twofold: the invariance property in the ratio of adjacent bin areas for polynomial tails and the fact that polynomial decay toward zero can be slow -- resulting in a large \( x_{n} \).

For general L\'evy processes, we can test whether the tail is exponential or polynomial.
For each candidate \( x_{\text{tail}} \), we compute approximations of the tail mass under both the exponential and polynomial decay assumptions, and compare these estimates with a high-precision numerical integration result (obtained via the Fortran library QUADPACK). 
 
In practice, for the intensities listed in Table~\ref{tab:crm} with typical parametrisations, this procedure is performed only once or twice. If one of the approximations is sufficiently close (\(\epsilon_{\mathrm{tail}}=10^{-5}\) by default) to the high-precision value, the corresponding \( x \) is accepted as \( x_{\text{tail}} \).

\subsubsection{Adaptive step for $x_0$ with fixed truncation length}
\label{section:lower}

Suppose that we truncate $\tilde{G}$ to the largest $N$ jumps. We want $x_0 < \eta_{\tilde\nu}(E_N)$ where $E_N$ is the $N$-th value of a unit rate Poisson process. 
We choose an initial grid $y_0, \dots, y_n$. If $y_0 \geq \eta_{\tilde\nu}(E_N)$. we find a sequence of additional grid points $z_{-1}=y_1 > z_0 = y_0 > z_1 > z_2 > \dots > z_K$ by choosing a step length $c > 0$ and using the following algorithm.

\begin{enumerate}
\item Set $i = 0$
\item Calculate 
$
b_{i} := \int_{z_{i-1}}^{z_{i}} \tilde{\nu}(z)\,dz
$
and 
$
m = \left\lceil \frac{E_N - \eta_{\tilde{\nu}}(z_{i})}{b_i} \right\rceil
$.
\item Set $z_i =y_0, z_{i+1}=c^{-1}z_i , \dots, z_{i + m} = c^{-m}z_{i}$.
\item If $\int_{z_{i + m}}^{y_n}
\tilde\nu(z) \,dz < E_N$, set $i = i + m$ and go to step 2.
\end{enumerate}
After the algorithm terminates, we set $(x_0, \dots, x_n) = (z_K, \dots, z_1, y_0, \dots, y_n)$, where $K$ is calculated by the steps above, 
so that the total mass over the extended grid is increased by at least \(E_N - \eta_{\tilde{\nu}}(x_0)\). In effect, adding these \(K\) points results in a new lower grid bound \(x_0\) that is shifted closer to zero.

This iterative search to find $x_0$ can be avoided if the L\'evy intensity close to zero is dominated by a polynomial term $x^{-\kappa}$, which is true for 
most L\'evy processes used in Bayesian nonparametrics  (all processes in Table \ref{tab:crm}).

If not explicitly stated by the user, we assess whether the L\'evy intensity \(\nu\) exhibits a dominant polynomial behaviour of the form \(x^{-\kappa}\) for \(x\) near zero and estimate the exponent \(\kappa\) following the procedure outlined in Section~\ref{sec:poly-tail}. In practice, we compute two estimates, \(\tilde{\kappa}_1\) and \(\tilde{\kappa}_2\), at two distinct small values \(x_{\mathrm{small}}^{(1)}\) and \(x_{\mathrm{small}}^{(2)}\) (for example, \(10^{-20}\) and \(10^{-30}\), respectively). If
\[
\left|\tilde{\kappa}_1 - \tilde{\kappa}_2\right| < \epsilon,
\]
for a prescribed tolerance \(\epsilon\), we assume that the intensity is dominated by the term \(x^{-\kappa}\).

Under this assumption, we employ the method used to choose $x_n$ when the L\'evy intensity has polynomially decaying tails. In the special case where \(\kappa = 1\), it follows that
\[
\frac{\displaystyle \int_{cx}^{c^2x} z^{-1}\,dz}{\displaystyle \int_{x}^{cx} z^{-1}\,dz} = 1.
\]
\section{Numerical Results}
\label{sec:numerical}
\subsection{Precision}
\label{sec:precision}
\begin{figure}[ht]
    \centering
    \begin{subfigure}{0.48\textwidth}
        \centering
        \includegraphics[width=0.8\textwidth]{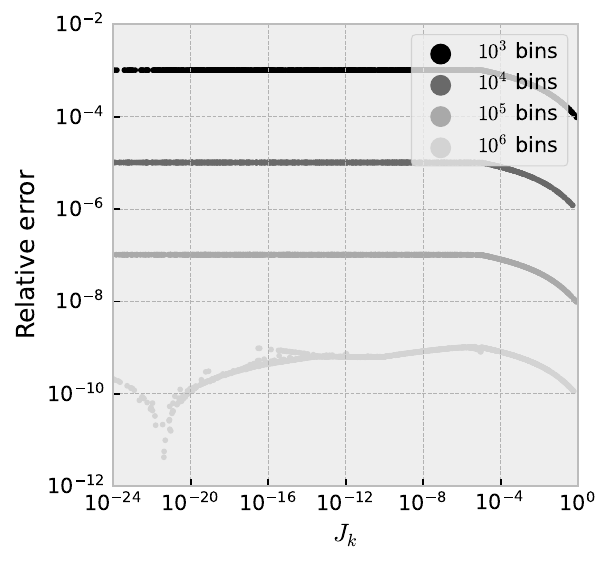}
        \caption{L\'evy intensity decomposition provided.  }
        
    \end{subfigure}
    \hfill
    \begin{subfigure}{0.48\textwidth}
    \centering
        \includegraphics[width=0.8\textwidth]{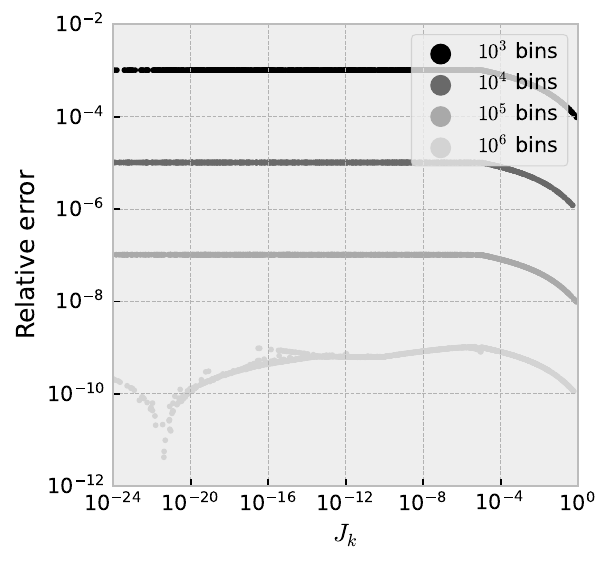}
        \caption{No L\'evy intensity decomposition provided.  }
        \label{fig:sub3} 
    \end{subfigure}
    \caption{Relative error of the jump sizes for various bin counts for the beta process with $M=1$ and $c=2$ ($x_{\text{thr}}=10^{-5}$). Providing L\'evy intensity decomposition leads to smaller error for the highest bin specification.}
    \label{fig:relative-error}
\end{figure}

\begin{figure}[ht]
    \centering
    \begin{subfigure}{0.48\textwidth}
    \centering
        \includegraphics[width=0.8\textwidth]{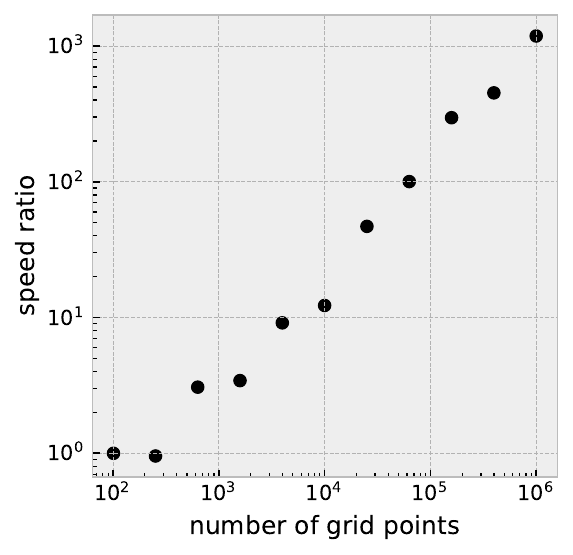}
        \caption{Speed ratio (vs 100 bin configuration) for simulation from the beta process $M=1$ and $c=2$ vs number of grid points (log-log scale). The speed is an approximately linear function of number of grid points.} 
        \label{fig:sub32}    
    
    \end{subfigure}
    \hfill
    \begin{subfigure}{0.48\textwidth}
        \centering
        \includegraphics[width=0.8\textwidth]{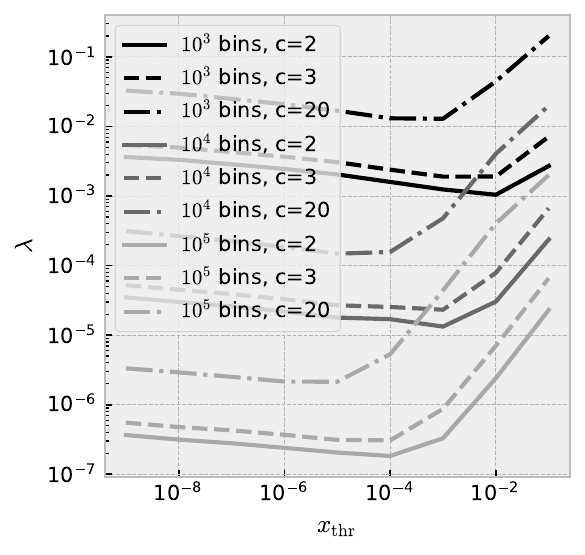}
        \caption{The mean number of the rejected points in the thinning correction versus number of grid points for the beta process with $M=1$ and various values for $c$ parameter. Analysis for truncation level $10^{-10}$.} 
        \label{fig:sub2}
    \end{subfigure}
    
    \caption{}
    \label{fig:test}
\end{figure}

\begin{figure}[ht]
    \centering
    \begin{subfigure}{0.48\textwidth}
        \centering
        \includegraphics[width=0.8\textwidth]{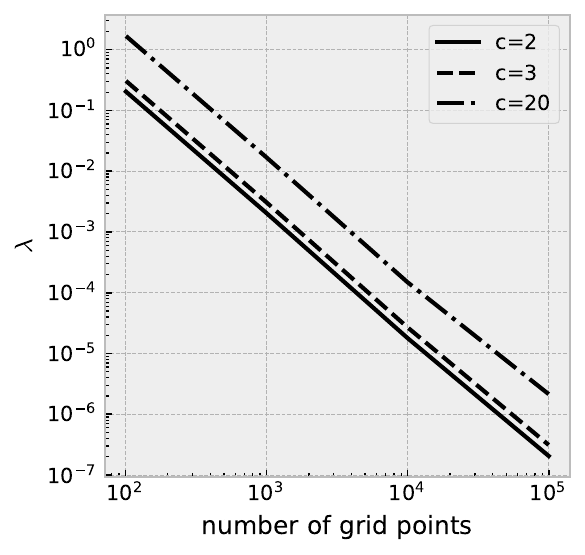}
        \caption{$x_{\text{thr}} = 10^{-5}$.}
        \label{fig:betapoi1}
    \end{subfigure}
    \hfill
    \begin{subfigure}{0.48\textwidth}
        \centering
        \includegraphics[width=0.8\textwidth]{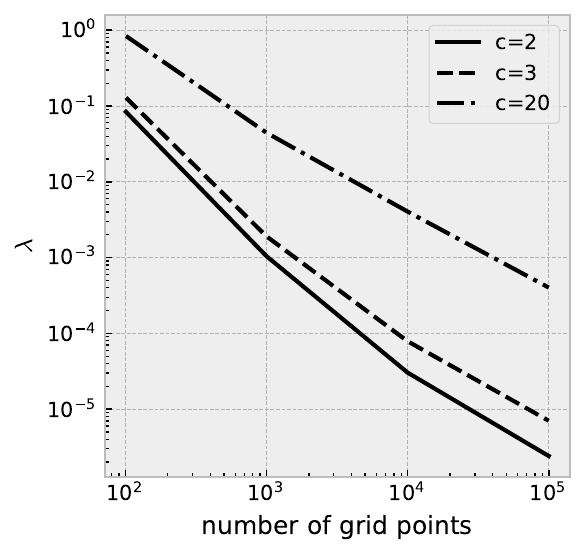}
        \caption{$x_{\text{thr}} = 10^{-2}$.} 
        \label{fig:betapoi2}
    \end{subfigure}
    \caption{The mean number of the rejected points in the thinning correction versus number of grid points for the beta process with $M=1$ and $c=2$ calculated with intensity decomposition provided. Results are same for no decomposition provided.}
    \label{fig:betapoi}
\end{figure}

The relative error of the jump sizes is dependent on the number of partition points. In Fig.\ \ref{fig:relative-error} we illustrate the relative error of the jump sizes, $\frac{|\tilde{J_k} - J_k|}{J_k}$, where $\tilde{J_k}$ is a jump size calculated using our algorithm without rejection and $J_k$ using Ferguson-Klass, as a function of the jump sizes and the number of partition points. As expected, the higher number of partition points has a lower relative error. The two-orders-of-magnitude decrease in relative error corresponds to one-order-of-magnitude more grid points as shown in Eq.\ \eqref{eq:relative-error}. For the bin configuration considered, $(c-1)^2$ yields the errors of magnitude $10^{-3}$, $10^{-5}$, $10^{-7}$ and $10^{-9}$ for $10^3$, $10^4$, $10^5$ and $10^6$ bins respectively. In this example, we used $x_{\mathrm{thr}}=10^{-5}$. The error due to the approximation of the decomposition of the intensity function (Section \ref{section:lower}) is negligible for the lower bin configuration but grows larger for the $10^6$ bin configuration.

The computation speed is an approximately linear function of the number of grid points, as illustrated in Fig.\ \ref{fig:sub3}, i.e., one order of magnitude increase in the number of grid points corresponds to one order of magnitude increase in the computation time as expected since the algorithm is linear in complexity. Results for other processes look very similar in terms of the magnitudes of the relative errors and speed ratios.

Fig.\ \ref{fig:sub2} shows the mean rejection in the thinning correction as a function of the number of bins and $x_{\mathrm{thr}}$. Parametrisation used for the beta process was  $M=1$, $c=\{2,3,20 \}$. Additionally, for the threshold,hold, we set $x_{\mathrm{thr}}= 10^{-5}$ in Fig.\ \ref{fig:betapoi1} and $x_{\mathrm{thr}}= 10^{-2}$ in Fig.\ \ref{fig:betapoi2}.
 
We show that the $x_{\text{thr}}$ parameter can be optimised based on the number of bins. This parameter gives lower rejection rate for the lower limit if the number of bins and for the upper limit if the number of the bins is small. Intuitively, this is because the trapezium rule has a smaller error when the subinterval width is very small (see Appendix \ref{appendix:integration} for analytical derivation). However, as seen in Fig.\ \ref{fig:sub2} the error rate needs to be balanced against the relative error metric for the optimal choice. 

\subsection{Performance for standard processes}
\label{sec:performance}
The proposed algorithm, due to its general purpose nature, is expected to be slower than the process-targeted algorithms such as the two-piece envelope for rejection sampling \citep{griffin2019two} or CHHB \citep{campbell2019truncated} or gamma process Lomax 
\citep{rosinski2001series} but faster than the universal Ferguson-Klass algorithm. The process-specific algorithms utilise either tabulated integrals which makes inverting CDF fast (gamma process) or have well-calibrated rejection envelopes \citep{griffin2019two, rosinski2001series}. For the performance analysis, we use the configuration with 1000 bins. In each simulation we draw 100 jump sizes.

All results presented will vary to some degree depending on the hardware used. We used Python 3.12 with Intel(R) Core(TM) i7-7700T CPU @ 2.90GHz  with 4 cores and 8 logical processors.

In the speed comparisons, our algorithm was initialized for every trial. In practice, if the random process does not change, for example, for a Monte Carlo-based simulation, the algorithm does not have to be re-initiated and the execution speed would be much higher (orders of magnitude) than reported here.

\begin{table}[ht]
\resizebox{\textwidth}{!}{%
\begin{tabular}{llll}
Process           & F-K speed ratio & Two-envelope speed ratio & CHHB/Lomax speed ratio \\ \hline \hline 
Beta              & 700-1500       & 0.06-0.20                  & 0.07-0.6                   \\
Stable-Beta       & 1000-4500         & 0.06-0.22                & 0.1-0.6                   \\
Gamma             & 15-22         & 0.004-0.005              & 0.007-0.008               \\
Generalized Gamma & 200-1200         & 0.012-018                 & 0.014-0.021                \\
\end{tabular}%
}
\caption{Speed comparison for our approach with the intensity decomposition provided vs other methods. The ratio is measured with the speed of our method in the denominator. We provide the speed ratio ranges which were obtained for runs with various parameters, namely: $M \in \{1,3,5,7,10 \}$, $c \in \{2,3,20\}$ and $\sigma \in \{0.1, 0.3, 0.9\}$}
\label{tab:mixed}
\end{table}

\begin{figure}[ht]
\centering
\includegraphics[width=0.9\textwidth,trim={0 0cm 0cm 0cm},clip]{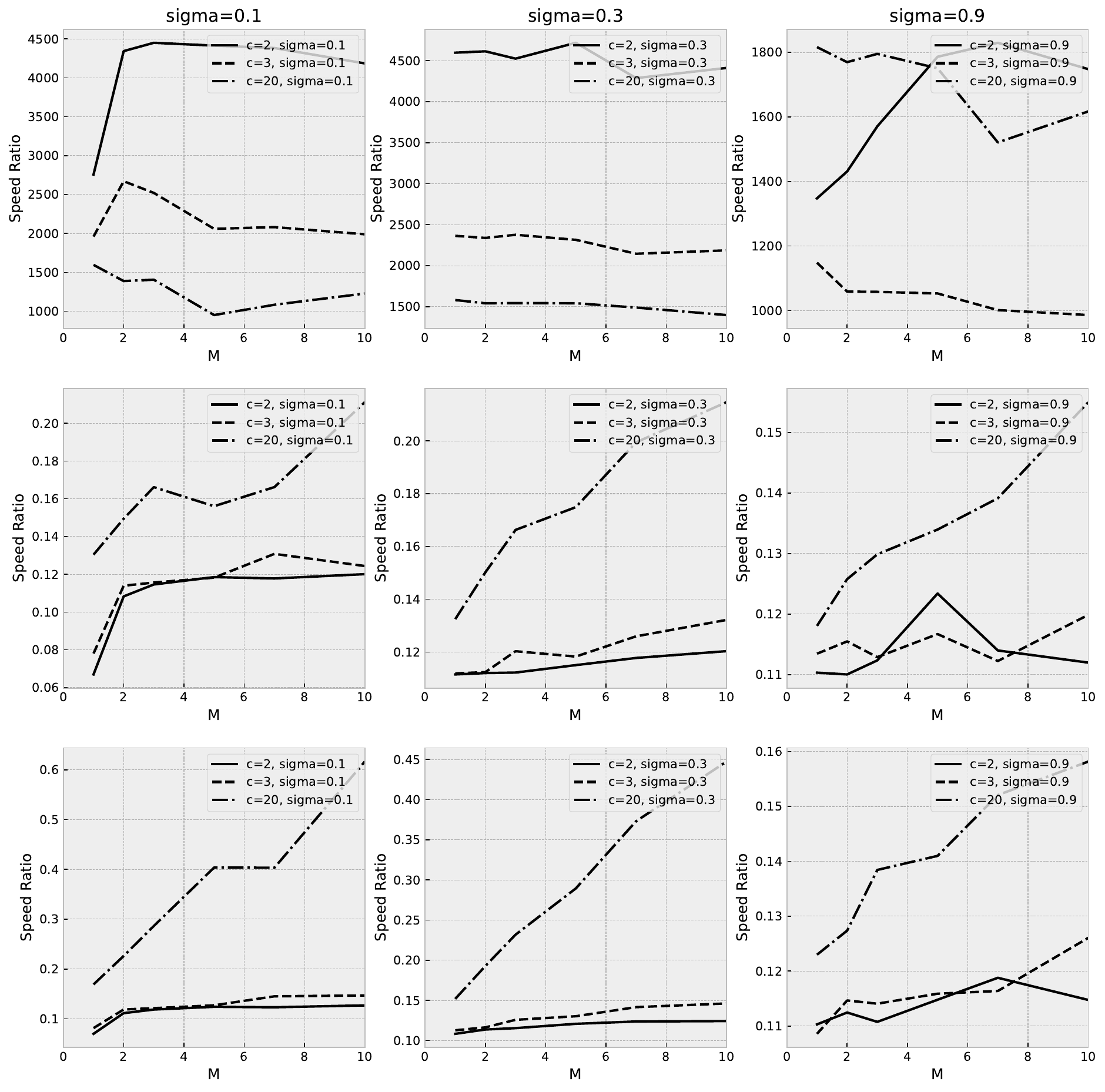}
\caption{Performance of the algorithm with 1000 bins with L\'evy intensity decomposition provided for a stable-Beta process versus Ferguson-Klass (top row),  two-piece envelope method (middle) and CHHB (bottom).}
\label{fig:stable-beta-perf}
\end{figure}

In Fig.\ \ref{fig:stable-beta-perf} we illustrate the performance comparison for the stable beta process. Table \ref{tab:mixed} summarises the performance metrics for the standard CRMs. For the stable beta process the proposed method with L\'evy intensity decomposition provided is around 1000-4500 times faster than the Ferguson-Klass algorithm and it is up to 10 times slower than the two-piece envelope or CHHB/Lomax algorithms. For the Beta process is 700-1500 times faster than Ferguson-Klass method specifically tailored to the the process and 5-17 times slower than the rejection based methods (two-piece envelope and CHHB/Lomax).

For the gamma process, our method is around 15 times faster than the Ferguson-Klass algorithm and about 15-100 times slower than the two-piece envelope method and the Lomax method \citep{rosinski2001series}.

\section{Examples}
\label{sec:examples}
In the Bayesian nonparametric literature, more intricate processes frequently emerge, either in calculating posteriors or during posterior computation.
 We provide two examples which justify the need for a general purpose algorithm for sampling from completely random measures. Further details about the Bayesian modelling are provided in Appendix~\ref{alg:back-examples}. In each case, we use 1000 bins and polynomial $f_i(x)$ close to zero with a known value of $\kappa$.

\subsection{Example 1} 
\label{sub:example1}

\cite{griffin2017compound} introduced
compound random measures (CoRMs) as a way to build a vector of $d$ correlated CRMs by perturbing the jumps of a one dimensional directing CRM with jump intensity $\nu^{\star}(z)$.  If  the CRMs have the same marginal L\'evy intensity, they show how
 $\nu^{\star}(z)$ can be derived from the distribution of the perturbations and the marginal jump intensity. A marginal beta process 
 (as given in Table 1)
 with  $\text{Beta}(\xi, 1)$ distributed perturbations gives, if $c > 1$, 
\begin{equation*}
    \nu^*(z) = M \,c \,z^{-1} \,(1-z)^{c-1} + \frac{M \,c \, (c-1)}{\xi} \,(1-z)^{c-2}.
\end{equation*}
This is the superposition of a  beta process with mass $M$ and a compound Poisson process  with intensity $M \,c \,/ \,\xi$ and jump distribution $\text{Beta}(1, c - 1)$. 
\begin{figure}[ht]
    \centering
    \includegraphics[width=\textwidth]{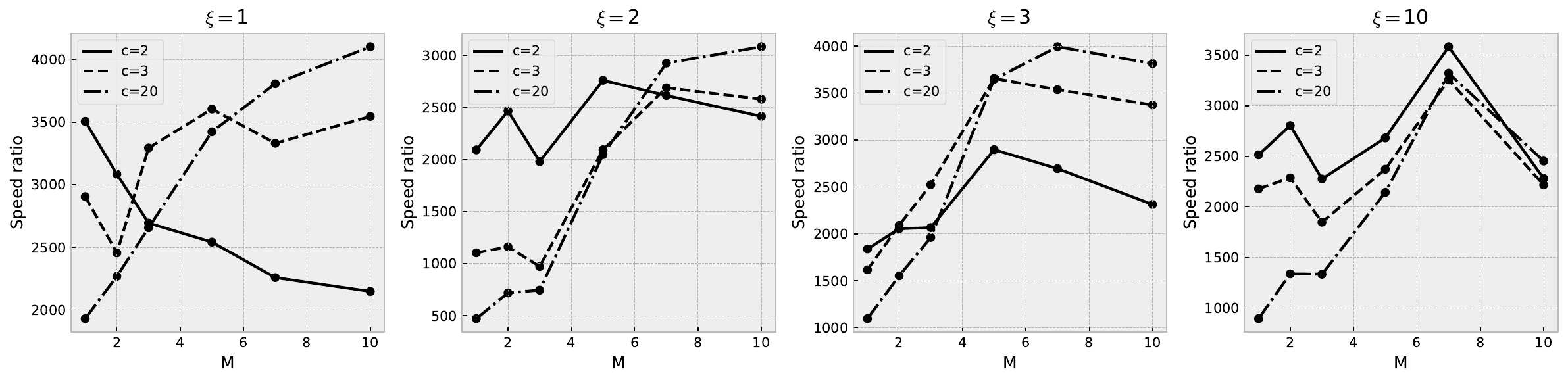}
    \caption{Example 1 -- The speed-up of our proposed algorithm over the Ferguson-Klass algorithm for all possible combinations of $c = 2, 3, 20$ and $\xi = 1, 2, 3, 10$ with $1 < M < 10$.}
    \label{fig:corm}
\end{figure}
We consider simulating this L\'evy intensity for all possible combinations of $c = 2, 3, 20$ and $\xi = 1, 2, 3, 10$ with $1 < M < 10$.
Figure~\ref{fig:corm} compares our algorithm with  the Ferguson-Klass algorithm.
The proposed method is two to three orders of magnitude faster than the Ferguson-Klass algorithm in this case.  The speed-up tends to be larger for larger $c$ and $M$.

\subsection{Example 2}
\label{sub:example2}

Occupancy models are commonly used in ecological and environmental statistics to model the observed presence of animal species in a study area
\citep[see][for a review]{devarajan2020multi}. Since animal species can be present but not observed, the model assumes an unknown probability of species presence and probability of detection. An animal is only observed if it is both present and detected. We use this example to
show how our method unlocks fast Bayesian nonparametric computations for  non-conjugate occupancy models. The model is similar to one presented in  \cite{masoero2022more}, who use cross-validation to fit the parameters by casting the model as a regression. We show how MCMC sampling can be performed in  non-conjugate model using the algorithm  proposed in this paper. 

We assume a simple model where there are an infinite number of species, which can be 
observed at $n$ sites on $K$ sampling occasions. 
 Let
 $\theta_i$ and $q_i$ be the probability of presence and probability of detection for the $i$-th species respectively and define $\theta = (\theta_1, \theta_2, \dots)$ and $q = (q_1, q_2, \dots)$.
  We assume that $\theta$ follows a beta process with L\'evy intensity $f(\theta)$ and the elements of $q$ are independent with prior distribution $p(q_i)$.
Suppose that $p^{\star}$ species are observed across the $n$ sites and $K$ sampling occasions and $Y_{i, j, k} = 1$ if the $i$-th of these species is observed at site $i$ on sampling occasion $k$ and $Y_{i, j, k} = 0$ otherwise. 
Define $I_{i, j} = \mbox{I}\left(\sum_{k=1}^K Y_{i, j, k} = 0\right)$.
Following
\cite{james2017},
the posterior is  a superposition of 
\begin{itemize}
\item Fixed points of discontinuity for the discovered species $i = 1, \dots, p^{\star}$ where
\begin{equation}
\label{eq:posterior-discovered}
p(\theta_i, q_i) \propto p(q_i) \, f(\theta_i)\, \prod_{j=1}^n \left[ \theta_i\,\prod_{k=1}^K q_i^{Y_{i, j, k}}\,(1 - q_i)^{1 - Y_{i, j, k}}  + 
(1 - \theta_i) \, I_{i, j}\right]
\end{equation}
\item A L\'evy process for the undiscovered species (for which $I_{i, 1} = \dots = I_{i, n} = 1$)  with tilted L\'evy intensity
\begin{equation}
\label{eq:posterior-undiscovered}
 p(q)\, f(\theta)\,
 \left[ \theta \, (1 - q)^K
 + 
(1 - \theta)
\right]^n
\end{equation}
\end{itemize}
The second part is a bivariate L\'evy process. 
One could expand the product using the binomial theorem to get a L\'evy intensity as a sum of a beta process and $n-1$ compound Poisson processes, but this could be time consuming to evaluate (c.f.\ Example \ref{sub:example1} for $n=2$) if $n$ is large and more complicated models (with covariates) would lead to even more complex expressions.
Our approach can be extended to this bivariate L\'evy process by  calculating the posteriors on a grid of points in the domain $[0,1]\times [0,1]$ where the grid in the dimension of $\theta$ is as specified in Section \ref{sec:methodology}, the grid in the $q$-dimension can be problem dependent, we keep it as uniform. The posterior density in the first part can be sampled 
 using the strip method described in \citet{Devr86}.


\cite{deng2019molecular} identified three main questions of interest in species sampling problems, namely, expected population frequency of a species with frequency $r \geq 1$, number of undiscovered species observed on $K_{\text{sub}}$ subsequent sampling occasions, and number of discovered species with frequency $r \geq 1$ that are expected to be observed on $K_{\text{sub}}$ subsequent sampling occasions. In this example we consider  the second question, that is, the distribution of number of undiscovered species observed on $K_{\text{sub}}$ subsequent sampling occasions. We also calculate the distribution of the number of undiscovered species observed $r$ times 
 on $K_{\text{sub}}$ subsequent sampling occasions. These are examples of an unseen species problem which was the focus of the recent study by \cite{balocchi2022bayesian}, where the authors used a species sampling model based on a Pitman-Yor prior.


\begin{figure}[ht!]
    \centering
    \begin{subfigure}{0.45\textwidth}
        \centering
        (a)\\
\includegraphics[width=0.8\textwidth]{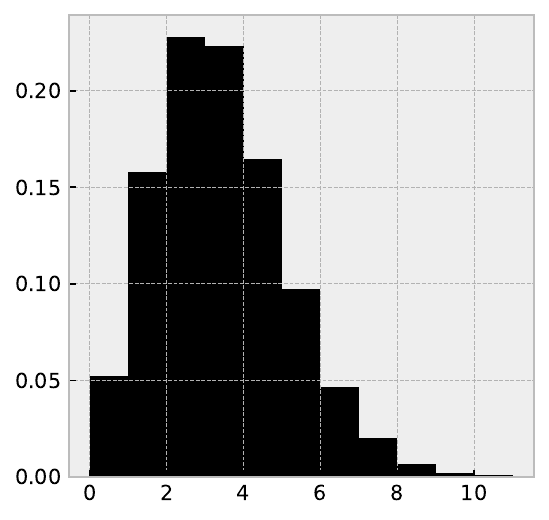}
    \end{subfigure}
    \hfill
    \begin{subfigure}{0.45\textwidth}
        \centering
        (b) \\
    \includegraphics[width=0.8\textwidth]{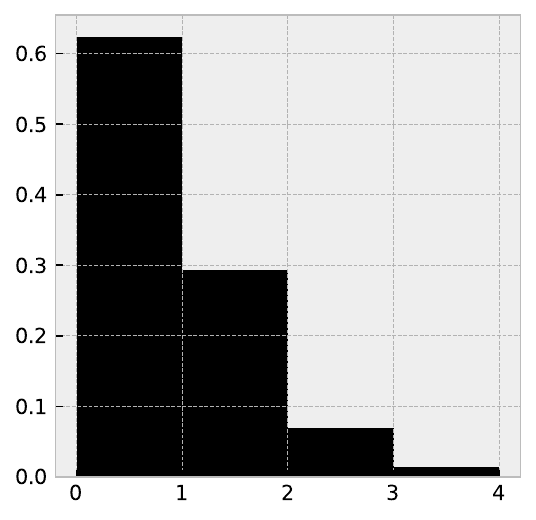} 
    \end{subfigure}
    \caption{Example 2 -- (a) 
    the distribution of the number of undiscovered species observed on 5 subsequent sampling occasions and (b)
    The distribution of the number of undiscovered species observed $r=2$ times on 5 subsequent sampling occasions.}
    \label{fig:occupancy-predictive}
\end{figure}

We consider 
the distributions of the number of undiscovered species and
the number of undiscovered species observed $r$ times 
found 
on $K_{sub}$ subsequent sampling occasions using Monte Carlo scheme. For each Monte Carlo sample, we generate
$(\theta^{\star}_1, r_1^{\star}), (\theta^{\star}_2, r_2^{\star}), (\theta^{\star}_3, r_3^{\star}), \dots$
from the L\'evy process in \eqref{eq:posterior-undiscovered}. We simulate, for $j = 1,\dots, n$ sites, 
$Z^{\star}_{i, j}$ where
\[
p(Z^{\star}_{i, j} = k) = \left\{\begin{array}{ll}
\theta_i^{\star} & \mbox{if }k = 1\\
0 & \mbox{otherwise}
\end{array}\right.
\]
 and $Y^{\star}_{i, j}\sim\mbox{Binomial}(5, r^{\star}_i)$  $\mbox{if }Z^{\star}_{i, j}=1$ and $Y^{\star}_{i, j} = 0$ if $Z^{\star}_{i, j} = 0$. The number of unobserved species is $\sum_i \sum_j \mbox{I}(Y_{i, j}^{\star} > 0)$ and 
 the number of unobserved species with frequency $r$ is 
  $\sum_i \sum_j \mbox{I}(Y_{i, j}^{\star} =  r)$. These provide draws from the corresponding distribution.

 We simulated synthetic data for 10 sampling sites with 5 initial sampling occasions
 using a beta process with $m=1$ and $c=2$, and the model parameters set to $\theta_1=0.6$, $\theta_2=0.06$, and $q=0.2$. We consider 5 subsequent sampling occasions and generated 10,000 samples from the distribution of unseen species 
(summarized in Fig.\ \ref{fig:occupancy-predictive}(a))
 and the distribution of unseen species with frequency $r = 2$ (summarized in
 Fig.\ \ref{fig:occupancy-predictive}(b)).
The number of jumps simulated from the L\'evy process was capped at 50. The total calculation time was around 5 minutes of which almost entire run time is spent on the repeated reevaluations of the predictive distribution based on the samples from the posteriors for the undiscovered species. Although setting grids and evaluating L\'evy intensity and posterior p.d.f take less than a second and is done only once the runtime cost is almost entirely consumed by evaluation of the conditional distribution $q|\theta$ and sampling of Bernoulli variables in the Monte Carlo step.
\section{Discussion}
In this paper we present a general-purpose approximation of the Ferguson-Klass algorithm to generate samples from a non-Gaussian L\'evy process. This approximation can be conceptualised as a type of rejection sampling algorithm that does not reject any samples due to a fine-grid envelope approximation. In that way it bridges the methods of \cite{rosinski2001series} and \cite{griffin2019two}. Our method is suitable for any non-Gaussian L\'evy process with a L\'evy intenstiy that can be expressed in an algebraic form. This is in contrast to the most well-known techniques, reviewed in \cite{campbell2019truncated}, which rely on approximations crafted to a handful of standard processes which historically limited the researchers to creation of conjugate models or models with severe computational limitations. Our method unlocks fast Bayesian nonparametric computations for non-conjugate models such as the species-sampling/occupancy model described in Example \ref{sub:example2}.

We show around 1000-fold improvement in speed over the Fergusson-Klass algorithm for well-known processes and around 500-4000 times speed-up for process mixtures as shown in Example \ref{sub:example1}. We analyse the behaviour of the error of our approximation and show that it is generally small or can be made small for certain selections of the approximation parameters in Section \ref{sec:precision}.

\bibliographystyle{chicago}                
\bibliography{references}

\newpage
\appendix

\section{Ferguson-Klass Algorithm}
\label{alg:ferguson-klass}
The algorithm below shows a sample implementation of the Ferguson-Klass algorithm \citep{ferguson1972representation}. The computational complexity can be calculated as $N \times \log[(b-a)/\mathrm{tol}] \times C_{\mathrm{integ}}$, where $K$ is the number of jumps, $\mathrm{tol}$ is the numerical tolerance in the root finding and the integration algorithms, and $C_{\mathrm{integ}}$ is the numerical complexity of the integration quadrature. The complexity of the integration step depends on the form of the L\'evy intensity and the exact algorithm used. E.g. if a simple trapezoidal method on a grid of $n$ nodes is used, the complexity would be $N \times n \times R$, where $R$ is the complexity of the root finding algorithm.

\begin{algorithm}[htbp]
\caption{\textsc{Ferguson--Klass Algorithm}}
\label{alg:ferguson_klass}
\begin{algorithmic}

\Require \\
\begin{itemize}
  \item Intensity function $\nu(x)$.
  \item Number of jumps $K$.
  \item Initial search interval $[\texttt{lower\_lim}, \texttt{upper\_lim}]$ for root-finding.
  \item Tolerance and method for the root-finding procedure (e.g.\ \texttt{Brent's}).
\end{itemize}

\Ensure
An array of roots, $\{\,x_1, x_2, \dots, x_n\}$, corresponding to each unit Poisson process arrival time $E_i \in T$.

\Statex

\Function{\textsc{FergusonKlass}}{$T$}
  \State \(\textit{jumpsExact} \gets \emptyset\) 
         \Comment{Initialize array of solutions (jumps) for each $t_i$}
  \For{\textbf{each} $k$ \textbf{in} $K$}
    \State Generate $T_k \sim \text{Exp}(1)$
    \State $E_k \leftarrow E_{k-1} + T_k$ \Comment{Arrival time of $k$-th Poisson event}
    \State \(\texttt{fun}(x) \gets E_k \;-\; \displaystyle\int_{x}^{\infty} p_x(z)\,dz\)
           \Comment{Define the function whose root we seek}
    \Statex

    \State \textbf{try}
    \State \quad \(r \gets \textsc{RootFind}\bigl(\texttt{fun}, 
                        [\texttt{lower\_lim}, \texttt{upper\_lim}]\bigr)\)
           \Comment{Use a root-finding method (e.g.\ Brent's method) to find $r$ s.t.\ $\texttt{fun}(r)=0$.}
    \State \quad \(\textit{jumpsExact}.\text{append}(r)\)
    \State \quad \(\texttt{upper\_lim} \gets r\)
    \State \quad \(\texttt{lower\_lim} \gets r/10^6\)

  \EndFor
  \State \Return \textit{jumpsExact}
\EndFunction

\end{algorithmic}
\end{algorithm}
\section{Algorithm}
\label{sec:algo}
Algorithm~\ref{alg:approx_method} summarizes our overall procedure for approximating the tail mass function and simulating jumps from a completely random measure (CRM) via adaptive numerical integration. The key advantage of our approach is that the numerical integration of tail masses is performed once for all arrival times of the unit Poisson process, thereby avoiding the repeated inversion of the tail mass function required by traditional methods (e.g., the Ferguson--Klass algorithm).

\begin{breakablealgorithm}
\label{alg:approx_method}
\begin{algorithmic}
\Require \\

\begin{itemize}
  \item L\'evy intensity \(\nu(x)\) defined on a domain \([a,b]\) (typically \(a=0\) and \(b=1\) or \(b=\infty\)); set the initial lower bound \(x_{0}\) (default \(10^{-10}\)).
  \item \textbf{Optional:} A factorisation \(\nu(x)=h(x)\,g(x)\) with \(h(x)=x^{-\kappa}\) (if available).
  \item \textbf{Optional:} Number of grid points \(n\) on \([x_{0},1]\) (default 1001) or geometric spacing factor \(c>1\) (default \(10^{10^{-2}}\)).
  \item \textbf{Optional:} Tolerance \(\epsilon\) for tail approximation (default \(10^{-10}\)).
  \item \textbf{Optional:} Threshold \(x_{\mathrm{thr}}\) separating the near-zero region from the rest (default \(10^{-2}\)).
\end{itemize}

\Ensure \\

\begin{itemize}
  \item An adaptive grid \(\{x_{0}, \dots,  x_{n}\}\) with computed bin masses \(\{b_i\}\).
  \item A piecewise interpolation function for the approximate tail mass \(\eta_{\tilde{\nu}}(x)=\int_x^{\infty}\tilde{\nu}(z)\,dz\).
  \item A realisation of the CRM \(G\) (either using the pure approximation or with optional rejection thinning).
\end{itemize}\\

\State \textbf{Step 1: Generate Poisson process arrival times.}
\State Generate $N$ unit Poisson process arrival times \(E_1, E_2,\dots,E_N\).\\


\State \textbf{Step 2: Approximate the right tail.}
\If{the domain of $\nu$ is bounded, e.g.\ $[0,b]$}
  \State $x_{n} \gets b$ \Comment{No infinite tail to approximate.}
\Else

    \For{$x \in \{10^0, 10^{0.5}, 10^1,\ldots \}$}

  \If{ If a special case is detected (Section \ref{section:tail}) \(x_{\text{tail}} \gets x\) and record the tail type.}
  \State  Find \(x_{n}\) so that
  \[
      \int_{x_{n}}^{\infty}\nu(x)\,dx < \epsilon,
  \]
  using the appropriate tail approximation.
  \ElsIf{$\eta_{\nu}(x) < \epsilon_{\text{large}}$}
  \[
  x_{n} \gets x
  \]
  \EndIf
  \EndFor
\EndIf\\

\State \textbf{Step 3: Build the primary grid.}
\State Create grid 
\[
   x_0 , x_1,\ldots,x_k = x_{\text{tail}},~
   x_{k+1},\ldots,x_n,
\]
\State Use geometric spacing on \([x_{0}, x_{\text{tail}}]\), i.e. \(\{ x_0, c x_0, c^2 x_0,\ldots\}\) and then either geometric or equally spaced points on \([x_{\text{tail}},x_{n}]\) as dictated by the tail type.\\

\State \textbf{Step 4: Compute bin masses.}
\State Calculate bin masses for $x \in [x_{\mathrm{thr}}, x_{n}]$.

  \For{each bin \(i\) with \(x \in [x_i, x_{i+1}]\) where \(x_i \ge x_{\mathrm{thr}}\)}
    \State 
    \[
      b_i \approx 0.5\,\Bigl[\nu(x_i)+\nu(x_{i+1})\Bigr](x_{i+1}-x_i).
    \]
\EndFor
  \State Calculate bin masses for $x \in [x_{0}, x_{\mathrm{thr}}]$.           
\If{a dominant polynomial behaviour near \(0\)  is present}
  \For{each bin \(i\) with \(x_i < x_{\mathrm{thr}}\)}
      \State Set \(g(x_i) \gets \nu(x_i) / x_i^{-\kappa}\) if the decomposition is not provided.
      \State
      \[
         b_i \approx g(x_i) \int_{x_i}^{x_{i+1}} z^{-\kappa}\,dz.
      \]
  \EndFor
\Else
  \For{each bin \(i\) with \(x_i < x_{\mathrm{thr}}\)}
      \State 
      \[
         b_i \approx 0.5\,\Bigl[\nu(x_i)+\nu(x_{i+1})\Bigr](x_{i+1}-x_i).
      \]
  \EndFor
\EndIf\\


\State \textbf{Step 5: Adaptive search for \(x_0\).}
\State Rename the grid used so far to \(\{y_0,\ldots,y_n \}\). Find a sequence of additional grid points $z_{-1}=y_1 > z_0 = y_0 > z_1 > z_2 > \dots > z_K$ by choosing a step length $c > 0$ and using the following algorithm.

\State Compute 
\[
   \eta_{\tilde{\nu}}(y_0) = \int_{y_0}^{y_n} \tilde{\nu}(z)\,dz = \sum_{i=0}^{N} b_i.
\]

\If{\(E_N > \eta_{\tilde{\nu}}(y_0)\)}
  \If{a dominant polynomial term is present near \(0\)}
    \State Compute 
    \[
       b_0 = \int_{y_0}^{c y_0} \tilde{\nu}(z)\,dz.
    \]
    \State Set
    \[
    K = \begin{cases}
         \left\lceil \log_{c^{\kappa-1}}\!\Biggl(1 - \dfrac{E_N}{b_0}\,(1-c^{\kappa-1})\Biggr) \right\rceil, & \text{if } \kappa\neq 1,\\[1mm]
         \left\lceil \dfrac{E_N -\eta_{\tilde{\nu}}(y_0)}{b_0} \right\rceil, & \text{if } \kappa=1.
    \end{cases}
    \]
    \State Set \( \{x_0,\ldots,x_n \} = \{z_m,\ldots,z_1,y_1,\ldots,y_n \}\).
  \Else
    \State set \(i=0\), \(m=0\)
    \While {$\int_{z_{i + m}}^{y_n}
\tilde\nu(z) \,dz < E_N$}
        \State Calculate 
$
b_{i} := \int_{z_{i-1}}^{z_{i}} \tilde{\nu}(z)\,dz.
$
and 
$
m = \left\lceil \frac{E_N - \eta_{\tilde{\nu}}(z_{i})}{b_i} \right\rceil
$.
    \State Set $z_i =y_0, z_{i+1}=c^{-1}z_i , \dots, z_{i + m} = c^{-m}z_{i}$.
    \State Set $i = i + m$
    \State Set \( \{x_0,\ldots,x_n \} = \{z_m,\ldots,z_1,y_1,\ldots,y_n \}\)
    \State Compute the bin masses for the additional bins as outlined in \textbf{Step 4}.
    \EndWhile
\EndIf
    \Else
     \State Set \( \{x_0,\ldots,x_n \} = \{y_1,\ldots,y_n \}\)

  \EndIf\\

\State \textbf{Step 6: Calculate tail mass for off-grid arrival times.}
\For{each arrival \(E_j\) that falls between two grid points \(x_i\) and \(x_{i+1}\)}
    \State Invert the integration function (trapezoidal or polynomial) to find $x: \eta_{\tilde{\nu}}(x)=E_j$. Alternatively, use linear interpolation as an approximation to this procedure
  \[
      \eta_{\tilde{\nu}}^{-1}(E_j) \approx \eta_{\tilde{\nu}}^{-1}(x_{i+1}) + \frac{\eta_{\tilde{\nu}}^{-1}(x_i)-\eta_{\tilde{\nu}}^{-1}(x_{i+1})}{x_i - x_{i+1}}\,(E_j - x_{i+1}).
  \]
\EndFor
\State \textbf{Note:} The linear interpolation introduces another approximation error but is computationally more efficient. It can be activated as an option.\\

\State \textbf{Step 7 (optional): Rejection thinning.}
\State Let \(\tilde{G}\) be the enveloping CRM with intensity \(\tilde{\nu}(x)\ge\nu(x)\).
\For{each simulated jump \(\tilde{J}_k\) from \(\tilde{G}\)}
   \State Accept \(\tilde{J}_k\) with probability \(\nu(\tilde{J}_k)/\tilde{\nu}(\tilde{J}_k)\); otherwise, reject it.
\EndFor\\

\State \textbf{Output:}
\begin{itemize}

      \item An adaptive grid $\{x_{0}, \ldots, x_n \}$ 
  with computed bin masses $\{b_{0},\ldots,b_{n-1}\}$. 
  \item A piecewise function for $\eta_{\tilde{\nu}}^{-1}(x)$
  \item A realisation of the CRM $G^{(N)}$ (with or without rejection thinning).
\end{itemize}

\end{algorithmic}
\end{breakablealgorithm}
\section{Integration}
\label{appendix:integration}



\noindent{\it Proof of Proposition 1}
We begin by applying the mean value theorem for integrals over \([x_i,x_{i+1}]\). There exists some \(c\in[x_i,x_{i+1}]\) such that
\[
I(i) = \int_{x_i}^{x_{i+1}} h(x)g(x)\,\mathrm{d}x = g(c) \int_{x_i}^{x_{i+1}} h(x)\,\mathrm{d}x.
\]
The \emph{polynomial rule} approximates the integral by taking the value of \(g\) at the left endpoint:
\[
I_{\mathrm{poly}}(i) = g(x_i) \int_{x_i}^{x_{i+1}} h(x)\,\mathrm{d}x.
\]
Thus, the error is
\[
e_{\mathrm{poly}}(i) = I(i) - I_{\mathrm{poly}}(i) = \bigl(g(c)-g(x_i)\bigr) \int_{x_i}^{x_{i+1}} h(x)\,\mathrm{d}x.
\]
Using a Taylor expansion of \(g\) around \(x_i\),
\[
g(c) = g(x_i) + g'(x_i)(c-x_i) + O((c-x_i)^2),
\]
and noting that \(c-x_i < \Delta x_i\) (with \(\Delta x_i=x_{i+1}-x_i\)), we deduce that, as $\Delta x_i \rightarrow 0$,
\[
e_{\mathrm{poly}}(i) = O\Bigl(g'(x_i)\Delta x_i \int_{x_i}^{x_{i+1}} h(x)\,\mathrm{d}x\Bigr).
\]
Noting that \[
\int_{x_i}^{x_{i+1}} h(x)\,\mathrm{d}x = O\Bigl(\max_{x\in[x_i, x_{i+1}]}\{h(x)\}\Delta x_i\Bigr),
\]
then
\[
e_{\mathrm{poly}}(i) = O\Bigl(g'(x_i)\,\max_{x\in[x_i, x_{i+1}]}\{h(x)\}(\Delta x_i)^2\Bigr).
\]
and, since $\max_{x\in[x_i, x_{i+1}]}\{g'(x_i) h(x)\} < \max_{x\in[x_i, x_{i+1}]}\{g'(x) h(x)\}$
\[
e_{\mathrm{poly}}(i) = O\Bigl(\max_{x\in[x_i, x_{i+1}]}\{g'(x)h(x)\}(\Delta x_i)^2\Bigr).
\]
Next, consider the \emph{trapezoidal rule} which approximates the integral by
\[
I_{\mathrm{trap}}(i) = \frac{\Delta x_i}{2}\Bigl[f(x_i) + f(x_{i+1})\Bigr] = \frac{\Delta x_i}{2}\Bigl[h(x_i)g(x_i)+h(x_{i+1})g(x_{i+1})\Bigr].
\]
Taylor series expansions of 
both \(h\) and \(g\) 
about \(x_i\)
yield
\[
h(x_{i+1}) = h(x_i) + h'(x_i)\Delta x_i + O((\Delta x_i)^2),
\]
and
\[
g(x_{i+1}) = g(x_i) + g'(x_i)\Delta x_i + O((\Delta x_i)^2).
\]
Multiplying these expansions, we have
\[
h(x_{i+1})g(x_{i+1}) = h(x_i)g(x_i) + \Delta x_i\Bigl[h(x_i)g'(x_i)+h'(x_i)g(x_i)\Bigr] + O((\Delta x_i)^2).
\]
Consequently, the trapezoidal approximation becomes
\[
I_{\mathrm{trap}}(i) = h(x_i)g(x_i)\Delta x_i + \frac{(\Delta x_i)^2}{2}\Bigl[h(x_i)g'(x_i)+h'(x_i)g(x_i)\Bigr] + O((\Delta x_i)^3).
\]
The properties of Riemann sums imply that 
\[
 -M_1\leq I(i) - h(x_i)g(x_i)\Delta x_i  \leq   M_1
\]
where $M_1 = \frac{1}{2}\max_{x\in [x_i, x_{i+1}]}\{\vert h(x) g'(x) + h'(x)  g(x)\vert\}(\Delta x_i)^2$.
The error is
$e_{\mathrm{trap}}(i) =  I(i) - I_{\mathrm{trap}}(i)$
then 
\[
- M_1  \leq 
e_{\mathrm{trap}}(i) + \frac{(\Delta x_i)^2}{2}\Bigl[h(x_i)g'(x_i)+h'(x_i)g(x_i)\Bigr] +
 O((\Delta x_i)^3)
\leq M_1.
\]
Clearly,
\[
M_1 + \frac{(\Delta x_i)^2}{2}\Bigl[h(x_i)g'(x_i)+h'(x_i)g(x_i)\Bigr] \leq (\Delta x_i)^2 \max_{x\in [x_i, x_{i+1}]} \Bigl[h(x)g'(x)+h'(x)g(x)\Bigr] 
\]
\[
M_1 - \frac{(\Delta x_i)^2}{2}\Bigl[h(x_i)g'(x_i)+h'(x_i)g(x_i)\Bigr] \leq \frac{1}{2}(\Delta x_i)^2 \max_{x\in [x_i, x_{i+1}]} \Bigl[h(x)g'(x)+h'(x)g(x)\Bigr].
\]
It follows that $e_{trap}(i) = O\left(\max_{x\in [x_i, x_{i+1}]} \Bigl[h(x)g'(x)+h'(x)g(x)\Bigr] \right)$. If $h'(x)g(x) > h(x) g'(x)$ then 
 $e_{trap}(i) = O\left(\max_{x\in [x_i, x_{i+1}]} \Bigl[h(x)g'(x)\Bigr] \right)$

Under the stated assumption that \(|h'(x)| \ge K\,|g'(x)|\) for large \(K\), the term \(h'(x_i)g(x_i)\) dominates. Therefore,
\[
e_{\mathrm{trap}}(i) = O\Bigl(g(x_i)h'(x_i)(\Delta x_i)^2\Bigr).
\]

Since \(K\gg1\) implies that \(|h'(x_i)| \gg |g'(x_i)|\) and assuming \(g(x_i)\) remains bounded away from zero, it follows that the leading error in the polynomial rule is smaller (by an order proportional to \(g'(x_i)\) instead of \(h'(x_i)\)) than the trapezoidal rule. That is,
\[
\lvert e_{\mathrm{poly}}(i)\rvert < \lvert e_{\mathrm{trap}}(i)\rvert.
\]
Plugging in \(h(x)=x^{-\kappa}\) yields the result.
\hfill$\Box$\\
\\

 

\noindent{\it Proof of Proposition 2}
Continuing from the previous proof, the error of the polynomial is
\[
e_{\mathrm{poly}}(i) = O\Bigl( g'(x_i)\,x_i^{-p}\,(\Delta x_i)^2\Bigr).
\]
and the trapezoidal method error is
\[
e_{\mathrm{trap}}(i)  
= O\Bigl( (\Delta x_i)^2 \Bigl[ p\,x_i^{-p-1}g(x_i) - x_i^{-p}g'(x_i) \Bigr]\Bigr).
\]
Since in the plateau region the decay of \(g(x)\) becomes significant, we generally have \(g(x_i)\) small and the ratio of the two errors is determined by
\[
\frac{\lvert e_{\mathrm{poly}}(i) \rvert}{\lvert e_{\mathrm{trap}}(i) \rvert} 
= O\left(\frac{g'(x_i)x_i^{-p}}{g(x_i)\,p\,x_i^{-p-1}}\right)
= O\left(\frac{x_i\,g'(x_i)}{p\,g(x_i)}\right).
\]
That is, one may write
\[
\frac{\lvert e_{\mathrm{poly}}(i) \rvert}{\lvert e_{\mathrm{trap}}(i) \rvert} 
= O\Bigl(\frac{x_i\,g'(x_i)}{g(x_i)}\Bigr).
\]
Now, if \(g(x)\) decays to 0 as \(x\to 1\) (or \(x\to\infty\)) in such a way that the ratio,
\(
\frac{g'(x)}{g(x)},
\)
remains sufficiently small relative to \(1/x\) (or more generally, if 
\(
\frac{x\,g'(x)}{g(x)} \ll 1 \quad \text{for } x\ge x_{\mathrm{thr}}),
\)
then the ratio above is less than one. This implies that
\[
\lvert e_{\mathrm{trap}}(i) \rvert < \lvert e_{\mathrm{poly}}(i) \rvert,
\]
for all \(x\ge x_{\mathrm{thr}}\) and sufficiently small \(\Delta x_i\).
\hfill$\Box$\\
\\

For example, if \(g(x)=(1-x)^{c-1}\) (as in the beta process) then 
\[
\frac{g'(x)}{g(x)} \approx -\frac{c-1}{1-x},
\]
and the condition becomes
\[
\frac{x}{1-x}(c-1)\ll 1,
\]
which is satisfied for \(x\) sufficiently close to 1. More generally, for any intensity function of the form 
\[
f(x)=x^{-p}g(x)
\]
with \(g(x)\to 0\) as \(x\to 1\) (or as \(x\to\infty\)), one may determine a threshold \(x_{\mathrm{thr}}\) such that for \(x\ge x_{\mathrm{thr}}\) the trapezoidal rule yields a smaller error than the polynomial rule.

We can also estimate the relative error of the jump sizes \(\frac{|J_k-\tilde{J}_k|}{J_k}\). For both the trapezoidal and the polynomial rule the error in each bin is of order \(e_i=O((\Delta x_i)^2)\). Let us denote \(\Delta \eta_k\) to be the error in the tail mass due to our approximation for $k$th jump size so that \(\eta_{\nu}(J_k)=E_k\) and $\eta_{\nu}(\tilde{J}_k)=E_k+\Delta \eta_k$. Then using the Taylor expansion and the fact that \(\eta'(J_k)=-\nu(J_k)\) we have
\[
    \eta_k(\tilde{J}_k) \approx \eta_k(J_k) + \eta'(\tilde{J}_k)(\tilde{J}_k-J_k)
\]
and
\[
    \Delta \eta_k \approx -\nu(J_k)(\tilde{J}_k-J_k).
\]
This gives
\[
    |J_k-\tilde{J}_k| \approx \frac{\Delta \eta_k}{\nu(J_k)}.
\]
Since the order of magnitude error for the tail mass is just a sum of individual $e_i$s we have that \(\Delta \eta_k=O((\Delta x_i)^2)\). From this, we deduce that, as $\Delta x_i \rightarrow 0$,
\[
    \frac{|J_k-\tilde{J}_k|}{J_k} \approx O\left(\frac{(\Delta x_i)^2)}{x_i}\right).
\]
For the proposed geometric grid, the ratio $\frac{(\Delta x_i)^2)}{x_i}=c-1$ which yields
\[
    \frac{|J_k-\tilde{J}_k|}{J_k} \approx O\left((c-1)^2\right).
\]

\subsection{Example: Beta process}
We show that the split between the polynomial and trapezoidal rules can be optimised either numerically or algebraically. For the beta process $M c x^{-1} (1-x)^{c-1}$ the assumption \(|h'(x)| \gg |g'(x)|\) does not hold close to 1. Therefore, we need to modify the expression for the error of the trapezoidal rule to $e_{\mathrm{trap}}=-\frac{(\Delta x)^3}{12}f''(\xi)$, where $x_1 \leq \xi \leq x_{i+1}$ \citep{atkinson1991introduction}. 
Note that this formula reduces to the derived error
\[
    f''(x) = h''(x) g(x) + 2 h'(x) g'(x) + h(x)g''(x).
\]
For \(h(x) = x^{-\kappa}\), \(h''(x)\) is the dominant term and we have that \(h''(x) \approx \frac{h'(x + \Delta x) - h'(x)}{\Delta x}\), so that \(h''(x) \propto \frac{h'(x)}{\Delta x}\) which recovers the original expression.

For the beta process 
\[
    e_{\mathrm{poly}}(i) = O(\frac{c(c-1)(1-x_i)^{c-2}}{x_i}(\Delta x_i)^2),
\]
\[
    e_{\mathrm{trap}}(i) = O(\frac{cx_i^{-3}(1-x_i)^{c-3}[2+2(c-3)x_i+(c-2)(c-3)x_i^2]}{12}(\Delta x_i)^3)
\]
and
\[
 \frac{e_{\mathrm{poly}}(i)}{e_{\mathrm{trap}}(i)} = O(\frac{x_i^2}{\Delta x_i}).
\]
The last ratio can become larger than 1 for very small intervals $\Delta x_i$.

For example, for \(M=1\), \(c=2\), \(|e_{\mathrm{poly}}(i)|\approx \frac{(\Delta x_i)^2}{x_i}\) and \(|e_{\mathrm{trap}}(i)|\approx \frac{(\Delta x_i)^3}{3}x_i^{-3}\) so that
\[
    \left| \frac{e_{\mathrm{poly}}(i)}{e_{\mathrm{trap}}(i)} \right| \approx \frac{x_i^2}{3 \Delta x_i},
\]
and since \(\Delta x = x (c-1)\),
\[
    \left| \frac{e_{\mathrm{poly}}}{e_{\mathrm{trap}}} \right| \approx \frac{x}{3 (c-1)}.
\]
For example, for \(10^3\) bins the optimal $x_{\mathrm{thr}}\approx 0.07$, for \(10^4\) bins $x_{\mathrm{thr}}\approx 0.007$ and for \(10^5\) bins $x_{\mathrm{thr}}\approx 0.0007$.

\section{Grid Spacing Selection}
\label{sec:grid-selection}

Let 
\[
x_i = c^i x_0,\quad i = 1,\dots,n,
\]
with constant \(c>1\). We wish to choose the grid spacing so that the error contribution in each bin of the trapezoidal rule is roughly the same. For the trapezoidal rule, the local integration error in the \(i\)th bin is
\[
e_i \approx -\frac{(\Delta x_i)^3}{12} \nu''(x), \quad \text{for } x_i \le x \le x_{i+1}.
\]
Since the error depends on the second derivative of the intensity function \(\nu(x)\), the optimal spacing depends on its local behaviour.

In many Bayesian nonparametric problems the L\'evy intensity usually of the form
\[
\nu(x)=h(x)g(x),
\]
where \(h\) is the dominant component following a power law (i.e. \(|h'(x)| > |g'(x)|\)). The following propositions cover these cases.\\
\\


\noindent{\it Proof of Proposition 3}
With \(h(x)=x^{-1}\), the intensity is dominated by a polynomial term so that, aside from the slowly varying factor \(g(x)\),
\[
\nu(x) \propto x^{-1}.
\]
Its second derivative is
\[
\nu''(x) \propto \frac{\mathrm{d}^2}{\mathrm{d}x^2} x^{-1} \propto x^{-3}.
\]
Thus, over a given bin,
\[
e_i \propto (\Delta x_i)^3\, x^{-3}.
\]
To equidistribute the error among bins, we require that
\[
(\Delta x_i)^3\, x^{-3} \approx \text{constant},
\]
which implies
\[
\Delta x_i \propto x.
\]
A grid with constant relative spacing (i.e. a geometric grid) exactly satisfies
\[
x_{i+1} - x_i \propto x_i.
\]
\hfill$\Box$\\
\\


\noindent{\it Proof of Proposition 4}
More generally, assume that
\[
\nu(x) \propto x^{-p}\quad \text{with } p>0, \, p \neq 1 ,
\]
so that
\[
\nu''(x) \propto x^{-p-2}.
\]
Then the local error per bin satisfies
\[
e_i \propto (\Delta x_i)^3\, x^{-p-2}.
\]
The goal is to equidistribute the error over the integration range. To this end, introduce a continuously varying parameter \(u\) on \([0,1]\) and write
\[
x = h(u),
\]
with \(h(0)=0\) and \(h(1)=1\). Uniform spacing in \(u\) (with spacing \(\Delta u\)) gives, to first order,
\[
\Delta x \approx h'(u)\Delta u.
\]
In the continuous limit, the total error is approximated by
\[
e \propto \int_0^1 \Bigl[h(u)^{-p-2}\,\bigl(h'(u)\bigr)^3\Bigr]\,\mathrm{d}u.
\]
We now choose \(h(u)\) to minimize \(e\) subject to the constraint
\[
\int_0^1 h'(u)\,\mathrm{d}u = h(1) - h(0) = 1.
\]
Introducing a Lagrange multiplier \(\lambda\), consider the variational functional
\[
L[h] = \int_0^1 \left[ h(u)^{-p-2}\,\bigl(h'(u)\bigr)^3 - \lambda\, h'(u) \right]\mathrm{d}u.
\]
Applying the Euler--Lagrange equation,
\[
\frac{\mathrm{d}}{\mathrm{d}u}\frac{\partial \mathcal{L}}{\partial h'(u)} - \frac{\partial \mathcal{L}}{\partial h(u)}=0,
\]
leads (after some algebra) to the condition
\[
h(u)^{-p-2}\,\bigl(h'(u)\bigr)^3 = \text{constant}.
\]
Since this product is independent of \(u\), the local error is equidistributed over \(u\). Rewriting the condition gives
\[
h'(u) \propto h(u)^{(p+2)/3}.
\]
Separating variables and integrating,
\[
\int h(u)^{-(p+2)/3}\,\mathrm{d}h(u) \propto \int \mathrm{d}u,
\]
which, upon solving and inverting the relationship, results in the grid spacing
\[
\Delta x \propto x^{(p+2)/3}.
\]
\hfill$\Box$

\subsection*{Remarks}

In this work, the geometric spacing (\(\Delta x \propto x\)) is adopted as the default because it optimally equidistributes the error when the dominant term in the L\'evy intensity is \(x^{-1}\) (Proposition~\ref{prop:geo-app}). In many Bayesian nonparametric settings, one typically encounters L\'evy intensities where
\[
1 \le \kappa < 2,
\]
so that
\[
1\le \frac{\kappa+2}{3} < \frac{4}{3}.
\]
This implies that even when \(h(x)=x^{-\kappa}\) with \(\kappa\neq 1\), the geometric grid provides a good approximation to the optimal spacing of Proposition~\ref{prop:geo2-app}. More sophisticated, bespoke spacing schemes may further reduce the integration error if necessary.
\section{Stable Beta - Precision Analysis}
\begin{figure}[ht]
    \centering
    \includegraphics[width=\textwidth]{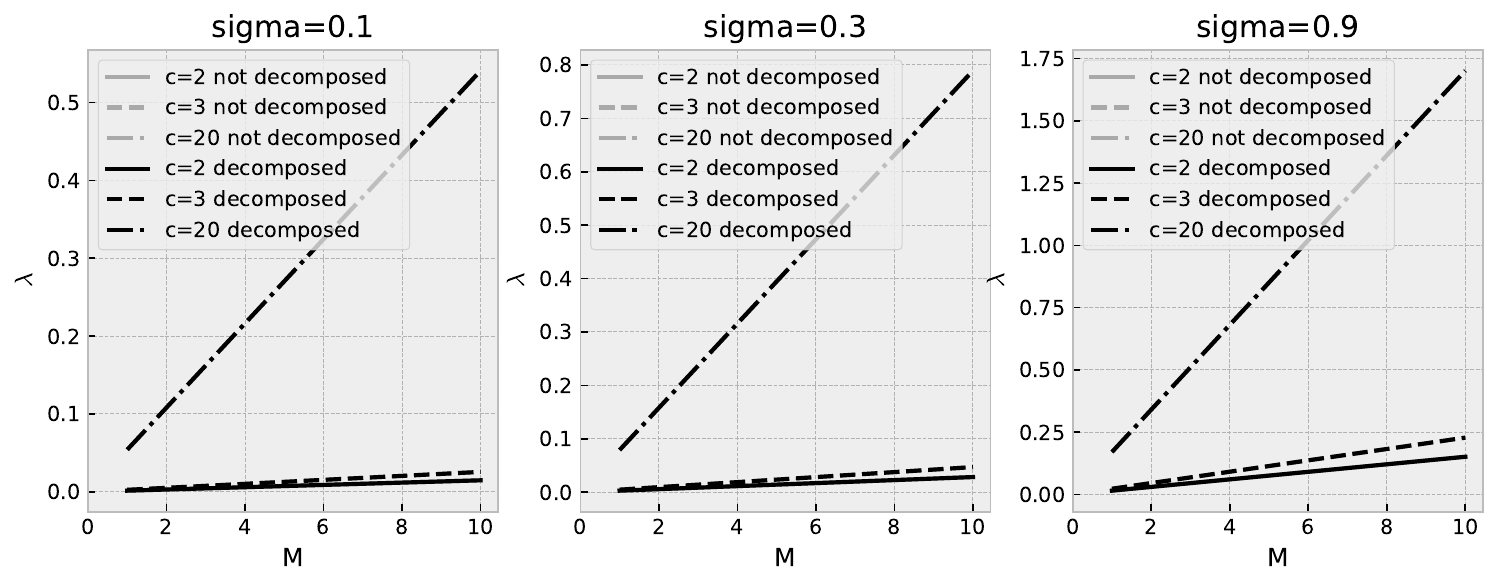}
    \caption{The mean of the Poisson distribution of the number of the points that should have been rejected for a stable beta process calculated using 1000 bins. The black lines correspond to the decomposed intensity and the blue lines correspond to the not decomposed intensity with $x_{\text{thr}}=10^{-2}$}
    \label{fig:stable-beta-rate}
\end{figure}

In this appendix we provide more detail on the approximation error analysis for the stable beta process. The calibration of the $x_{\text{thr}}$ parameter plays greater role for processes with fatter tail such as stable beta process. In Fig.\ \ref{fig:stable-beta-rate} we show that the mean of the Poisson distribution of the number of the points that should have been rejected is similar for the decomposed and not decomposed intensities. In this setting we integrated for $x$ between $10^{-10}$ and $1$ for the default settings for $x_{\text{lower}}$.

\begin{figure}[ht]
    \centering
    \begin{subfigure}{0.32\textwidth}
        \centering
        \includegraphics[width=\textwidth]{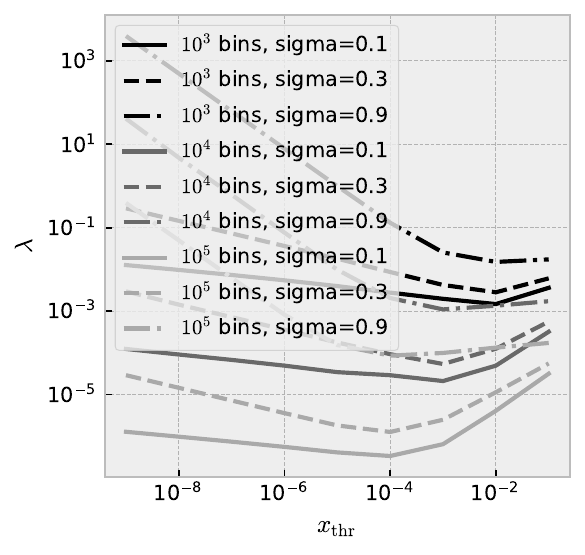}
        \caption{$M=1$, $c=2$.}
        \label{fig:stable-beta-1}
    \end{subfigure}
    \hfill
    \begin{subfigure}{0.32\textwidth}
        \centering
        \includegraphics[width=\textwidth]{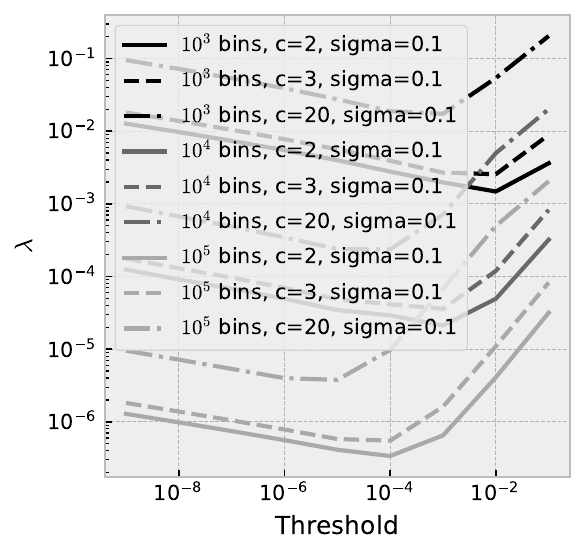}
        \caption{$M=1$, $\sigma=0.1$.} 
        \label{fig:stable-beta-12}
    \end{subfigure}
        \hfill
    \begin{subfigure}{0.32\textwidth}
        \centering
        \includegraphics[width=\textwidth]{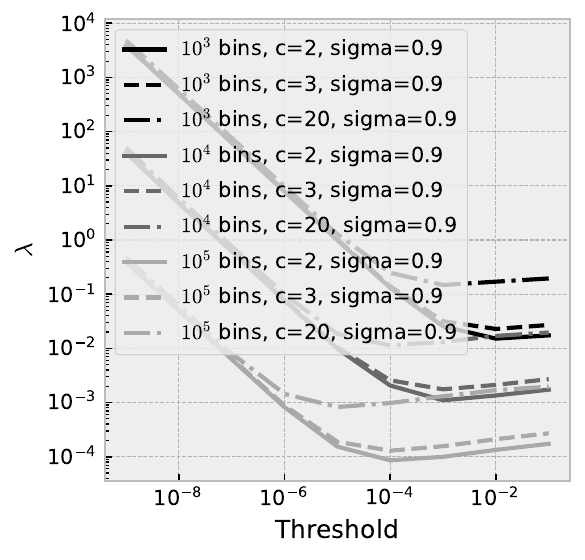}
        \caption{$M=1$, $\sigma=0.9$.} 
        \label{fig:stable-beta-3}
    \end{subfigure}
    \caption{The mean of the Poisson distribution of the number of the points that should have been rejected versus number of grid points for the beta process with $M=1$.}
    \label{fig:stable-beta-thr}
\end{figure}

In Fig.\ \ref{fig:stable-beta-thr} we show how the parameter $x_{\text{thr}}$ can be optimised with respect to the number of bins and the number bins and the stable beta parameters. For the approximation with a smaller number of bins $x_{\text{thr}}$ should be set closer to $10^{-2}$ and for the approximation with a lager number of bins $x_{\text{thr}}$ should be lowered.

\section{Background on Examples}
\label{alg:back-examples}

\subsection{Example 1}

The essence of the idea of compound random measures is to define $d$ correlated measures by perturbing the jumps of a one dimensional CRM. That is, if $\mu_j$ represents the $j$th random measure in the vector of random measures  $\mu_1,\ldots,\mu_d$ then
\begin{align*}
    \mu_j &= \sum_{i=1}^{\infty} m_{ji}\, J_i\, \delta_{x_i} \quad m_{1i},\ldots m_{di} \overset{\text{iid}}{\sim} h,\\
    \eta &= \sum_{i=1}^{\infty} J_i\, \delta_{x_i},
\end{align*}
where $m_{ij}$ are called scores and $\eta$ is a completely random measure with L\'evy intensity $\nu^* (\d s) \, \alpha(\d x)$. Therefore, CoRMs are fully characterised by the score distribution $h$ and the directing  L\'evy intensity $\nu^*$ or the directing  L\'evy intensity $\nu$ of the marginal process, since \cite{griffin2017compound} show that each marginal process has the same L\'evy intensity of the form
\begin{equation*}
    \nu_j(\d s) = \nu(\d s) = \int z^{-1} \, f(s/z) \, \d s \, \nu^*(\d z).
\end{equation*}

It is of interest to either calculate the directing L\'evy process based on the marginal L\'evy process and the score distribution or vice versa.  \cite{griffin2017compound} present some special cases where it's possible to do these calculations analytically, however, in general, the L\'evy process occurring in the calculations might be non-standard and intractable analytically. 

An example of a calculation for CoRMs shown in \citet{griffin2017compound} is a result for $\text{Beta}(\xi, 1)$ distributed scores together with the beta process.

Let the scores be $\text{Beta}(\xi, 1)$ distributed:
\begin{equation*}
    f(x) = \xi\, x^{\xi-1} \quad \xi>0, \quad 0<x<1.
\end{equation*}
If the L\'evy intensity of the jumps be a Beta process $\nu(s) = M c x^{-1} (1-x)^{c-1}$, $0<s<1$ and $c>1$, where $M$ is the mass parameter, then $\nu^*(z)$ is the solution of the integral equation
\begin{equation*}
    \nu(x) = \int_x^1 f(x/z) \, z^{-1} \, \nu^{*}(z) \,\d z, \quad 0<x<1.
\end{equation*}
The application of the fundamental theorem of calculus gives
\begin{equation*}
    \nu^*(z) = M \,c \,z^{-1} \,(1-z)^{c-1} + \frac{M \,c \, (c-1)}{\xi} \,(1-z)^{c-2}.
\end{equation*}
It is the sum of the L\'evy intensity of the original beta process with mass $M$ and a compound Poisson process (if $c>1$) with intensity $M \,c \,/ \,\xi$ and jump distribution $\text{Beta}(1, c - 1)$. Note that for $c=1$ the above only holds when $\xi=1$.

\subsection{Example 2}

In this example we show how our method unlocks fast Bayesian nonparametric computations for alternative formulations of occupancy models commonly used in ecological and environmental statistics. Specifically, we consider a non-conjugate formulation of the occupancy model used for the optimal allocation of resources in genomic variant discovery as presented in \cite{masoero2022more}. The authors do not perform inference via Markov Chain Monte Carlo (MCMC) but proceed to fit the parameters by casting the model as a regression with cross-validation and avoid the non-conjugate setup. In this example we show how MCMC sampling could be performed once we allow for non-conjugate model utilising fast sampling proposed in this paper. 

Similar occupancy models were also proposed for uses in ecology for modelling of the presence of species at the surveyed sites \citep{griffin2020modelling}. A review of such models can be found in \cite{devarajan2020multi}.

Suppose that we observed the presence/absence of multiple species at multiple sites or in multiple experiments (we will just refer to these as sites in the rest of this example). Let $Y_{i, j}$ denote the observed presence of the $i$-th species at the $j$-th site. A simple model is
\begin{equation}
p(Y_{i,j}=1\mid \theta_i) =\theta_i,\quad 
p(Y_{i,j}=1\mid \theta_i) = 1- \theta_i
\label{simple_model}
\end{equation}
where $\theta_i$ is the probability of observing the $i$-th species. This corresponds to the model in \cite{masoero2022more}
with $\theta_1, \theta_2, \dots $  given a particular nonparametric prior. In practice, a species may be present but not observed as present due to sequencing errors in experiments or non-detection of a species at a site. The simple model in \eqref{simple_model} can be extended by defining the actual presence of $i$-th species at the $j$-th site as $Z_{i, j}$.
\cite{masoero2022more}  assume that sampling effort differs between two experiments and is recorded as $s_j$ for the $j$-th experiment ($j = 1,2$). 
For $i=1,2, \dots$ and $s_j = 1, 2$, 
the model is extended  to
\begin{align*}
p(Z_{i,j} = 1\mid \theta_i)  = \theta_i, &\quad
p(Z_{i,j} = 0\mid \theta_i)  = 1 - \theta_i,
\\
p(Y_{i, j} = 1 \mid Z_{i, j} = 1, \lambda, s_j) = 1 - \exp\{- \lambda s_j\},&\quad 
p(Y_{i, j} = 0 \mid Z_{i, j} = 1, \lambda, s_j) = \exp\{- \lambda s_j\}
\\
  p(Y_{i, j} = 1 \mid Z_{i, j} = 0) = 0,
 &\quad 
 p(Y_{i, j} = 0 \mid Z_{i, j} = 0) = 1.
\end{align*}
They show how this model could be analysed analytically but we could consider extending the model in many different directions to make it more realistic. For example, we could allow for a species specific detection rate by changing $\lambda$ to $\lambda_i$ and assuming that $\lambda_i$ follows a beta distribution.
These changes can make the analysis more challenging.


We consider an extension to many sites and multiple sampling occasions at each site, which is natural in ecological surveys. 
Assuming that sampling effort is same at all sites and sampling occasions, we use the model, for 
 $i=1,\dots, p$, $j = 1,\dots,n$ and $k = 1,\dots, K$,
\begin{align*}
p(Z_{i, j}=1\mid \theta_i) =\theta_i, &\quad 
p(Z_{i, j}=0\mid \theta_i) =1-\theta_i
\\
p(Y_{i, j, k} = 1 \mid Z_{i, j} = 1, q_i) = q_i, 
&\quad p(Y_{i, j, k} = 1 \mid Z_{i, j} = 1, q_i) = 1- q_i
\\
p(Y_{i, j, k} = 1 \mid Z_{i, j} = 0) = 0, &\quad 
p(Y_{i, j, k} = 0 \mid Z_{i, j} = 0) = 1.
\end{align*}
where $\theta_i$ is the probability of the existence of the species $i$ and $q_i$ is the probability of detecting the species $i$ on any sampling occasion conditional on the presence of this species at the  site.
 By the law of total probability, the probability of the observed data at site $j$ for the $i$-th species is
\begin{align*}
p(Y_{i, j, 1}, \dots, Y_{i, j, K})
 =  \theta_i\, \prod_{k=1}^K q_i^{Y_{i, j, k}}\,(1 - q_i)^{1 - Y_{i, j, k}}  + 
(1 - \theta_i) \, \mbox{I}(Y_{i, j, 1} = \dots = Y_{i, j, K} = 0).
\end{align*}
The probability for the $i$-th species at all sites is 
\[
p(Y_{i, 1, 1}, \dots, Y_{i, 1, K}, \dots, 
Y_{i, n, 1}, \dots, Y_{i, n, K}) = \prod_{j=1}^n \left[ \theta_i\,\prod_{k=1}^K q_i^{Y_{i, j, k}}\,(1 - q_i)^{1 - Y_{i, j, k}}  + 
(1 - \theta_i) \, I_{i, j}\right]\\
\]
where $I_{i, j} = \mbox{I}\left(\sum_{k=1}^K Y_{i, j, k} = 0\right)$.

A species sampling model can be built by assuming that $\theta_1, \theta_2, \theta_3, \dots$ follow a beta process or three-parameter beta process and $q_i$ is given a parametric prior distribution. Let $p(q_i)$ be the prior density of $q_i$ and $f(\theta_i)$ be the L\'evy intensity of the process prior for $\theta_i$.

We refer to a species observed during the original $K$ sampling occasions as ``discovered''
(for which $\sum_{j, k} Y_{i, j, k} > 0$)
and the other species as ``undiscovered''. 
Assuming that there are  $p^{\star}$ 
discovered species then, following
\cite{james2017},
the posterior distribution of $(\theta_i, q_i)$ would be a superposition of 
\begin{itemize}
\item Fixed points of discontinuity for the discovered species $i = 1, \dots, p^{\star}$ where
\begin{equation}
\label{eq:posterior-discovered_a}
p(\theta_i, q_i) \propto p(q_i) \, f(\theta_i)\, \prod_{j=1}^n \left[ \theta_i\,\prod_{k=1}^K q_i^{Y_{i, j, k}}\,(1 - q_i)^{1 - Y_{i, j, k}}  + 
(1 - \theta_i) \, I_{i, j}\right]
\end{equation}
\item A L\'evy process for the undiscovered species (for which $I_{i, 1} = \dots = I_{i, n} = 1$)  with titled L\'evy intensity
\begin{equation}
\label{eq:posterior-undiscovered_a}
 p(q)\, f(\theta)\,
 \left[ \theta \, (1 - q)^K
 + 
(1 - \theta)
\right]^n
\end{equation}
{\it i.e.} the likelihood when $\sum_j \sum_k Y_{i, j, k}   = 0$ times the prior intensity. 
\end{itemize}

\end{document}